\documentclass[twocolumn,showpacs,preprintnumbers,amsmath,amssymb]{revtex4}
\usepackage{epsfig}
\usepackage{amsmath}

\begin{document}

\title{Activation Confinement Inside Complex Networks Communities}

\author{Luciano da Fontoura Costa}
\affiliation{Institute of Physics at S\~ao Carlos, University of
S\~ao Paulo, P.O. Box 369, S\~ao Carlos, S\~ao Paulo, 13560-970 Brazil}

\date{4rd Feb 2008}

\begin{abstract}
In this work it is described how to enhance and generalize the
equivalent model (arXiv:0802.0421) of integrate-and-fire dynamics in
order to treat any complex neuronal networks, especially those
exibiting modular structure.  It has been shown that, though involving
only a handful of equivalent neurons, the modular equivalent model was
capable of providing impressive predictions about the non-linear
integrate-and-fire dynamics in two hybrid modular networks. The
reported approach has also allowed the identification of the causes of
transient spiking confinement within the network communities, which
correspond to the fact that the little activation sent from the source
community to the others implies in long times for reaching the
nearly-simultaneous activation of the concentric levels at the other
communities and respective avalanches.  Several other insights are
reported in this work, including the smoothing of the spiking
functions, the consideration of intra-ring connections and its
effects, as well as the identification of how the weights in the
equivalent model change for different source nodes.  This work has
paved the way for a number of promising developments, which are
identified and discussed.  Preliminary results are also described
which reveal waves induced by the integrate-and-fire dynamics along
the steady-state regime.
\end{abstract}

\maketitle

\vspace{0.5cm}
\emph{`Eutropia is not one, but all these cities together; only
one is inhabited at a time, the others are empty...' 
(Invisible Cities, I. Calvino)}

\section{Introduction} 

Complex networks (e.g.~\cite{Albert_Barab:2002, Dorogov_Mendes:2002,
Newman:2003, Boccaletti:2006, Costa_surv:2007}) are interesting and
useful because they exhibit structured topology departing
significantly from uniformly random connectivity.  While the famous
scale free networks (e.g.~\cite{Albert_Barab:2002}) present degree
heterogeneity enhancing the probability of hubs, several real-world
networks exhibit modular organization, i.e. they include communities.
Informally speaking, communities are portions of the network (nodes
and respective edges) which are more intensely connected one another
than with the remainder of the network
e.g.~\cite{Girvan:2002,Zhou:2003, Newman:2004, Hopcroft:2004,
Radicchi:2004, Capocci:2005,Latapy:2005, Pons_comm:2005,
Eisler:2005,Yang:2005,Arenas_Vicente:2006, Boccaletti:2007,
Arenas:2008}.  Networks including communities are very important in
theory and practice exactly because of their modularity, which implies
strong effects on the respective topological and dynamical features.
For instance, the edges implementing the connections between two
communities tend to receive particularly intense activation during
dynamics (e.g. information transfer or transportation), corresponding
to functional bottlenecks.  Interestingly, several of the real-world
networks present modular organization.  For instance, the urban/land
transportation system is underlain by a complex network including the
communities (towns and cities) and intercommunity connections (roads).
Other especially important systems exhibiting modularity includes the
brain, collaborations, and economic systems, amongst many other
examples.

Because of the special importance of modularity for the structure and
dynamics of complex networks, much attention has been focused on the
problem of identifying the respective communities
(e.g.~\cite{Girvan:2002, Zhou:2003, Newman:2004, Hopcroft:2004,
Radicchi:2004, Capocci:2005, Latapy:2005, Pons_comm:2005,
Eisler:2005,Yang:2005, Arenas_Vicente:2006, Boccaletti:2007,
Arenas:2008}).  More recently, investigations targeting transient
non-linear dynamics in integrate-and-fire complex neuronal networks
indicated that the transient activation of neurons by a source placed
at one of the nodes tend to be confined during a transient period in
the community to which the source node belongs~\cite{Costa_begin:2008,
Costa_activ:2008}.  These results have important implications both for
neuroscience and complex network research.  In the former case, the
fact that topological localization (i.e. communities) tend to localize
neuronal activation in time, seems to be compatible with the fact
that, normally, the cognitive tasks have to be performed within
limited intervals of time inside specific functional modules
(e.g.~\cite{Amit:1992, Zeki:1999, Hubel:2005}).  Most other
traditional types of dynamics (e.g. traditional random walks or
self-avoiding random walks) are unable to yield such a confinement of
the dynamics inside the communities.  So, it could be possible that
the non-linear combination of the integration and fire elements in
neuronal cells are a necessary requisite for time-space
compartmentalization of information processing in neuronal systems.

It is interesting to observe that many other real-world systems and
problems also involve similar threshold-based dynamics, including
disease spreading, transportation, production (where the output goods
depends on the arrival of all parts), information processing and
computing, amongst many others.  Therefore, studies of the
relationship between the integrate-and-fire dynamics and the
topological communities have potential for applications in all these
real-world problems. Several are the implications of such studies also
for complex network research.  In particular, this issue lies at
the very heart of the important structure-and-dynamics paradigm, in
which relationships between the topological and functional properties
of networks are sought and investigated. 

In a work~\cite{Costa_nrn:2008} preceding the identification of the
spiking confinement in integrate-and-fire complex neuronal
networks~\cite{Costa_begin:2008, Costa_activ:2008}, another remarkable
effect had been observed.  More specifically, the activation, through
a single node, of several types of integrate-and-fire complex systems
was found to lead to \emph{avalanches} of activation and spikings.
Such a phenomenon has been observed for the Erd\H{o}s-R\'enyi and
Barab\'asi-Albert models, as well as for two types of knitted networks
(the path-regular and path-transformed BA
structures~\cite{Costa_comp:2007}).  However, no avalanches have been
observed for geographical integrate-and-fire networks~\footnote{The
phenomenon seems to occur sporadically in Watts-Strogatz networks.}.
This interesting phenomenon was later~\cite{Costa_equiv:2008} found to
be strongly defined by the hierarchical (or concentrical)
organization~\cite{Costa:2004, Costa_NJP:2007, Costa_JSP:2006,
Costa_EPJB:2005} of the respective complex networks.  More
specifically, the number of nodes in each of the concentric levels,
defined by the selection of the source node as a topological
reference, play a fundamental role in defining not only the timing of
the activation of each concentric level, but also the instant and
intensity of the main avalanches.  In particular, the main avalanche
tends to be produced by the almost simultaneous spiking of all neurons
in the concentric level containing the largest number of
neurons~\cite{Costa_equiv:2008}.

Assuming that the networks exhibit a reasonable level of degree
regularity, such a strong relationship between the concentric
structure and activation allowed a simple but effective equivalent
model to be defined which captures the intrinsic dynamics of avalanche
formation.  By using this equivalent model, which consists simply of a
chain integrate-and-fire network with weights and varying thresholds,
it is possible to make reasonably accurate predictions about the
intensity and time of the avalanches even for non-regular degree
networks such as the Barab\'asi-Albert model and the real-world
\emph{C. elegans} neuronal network.  

\begin{figure*}[htb]
  \vspace{0.3cm} 
  \begin{center}
  \includegraphics[width=0.3\linewidth]{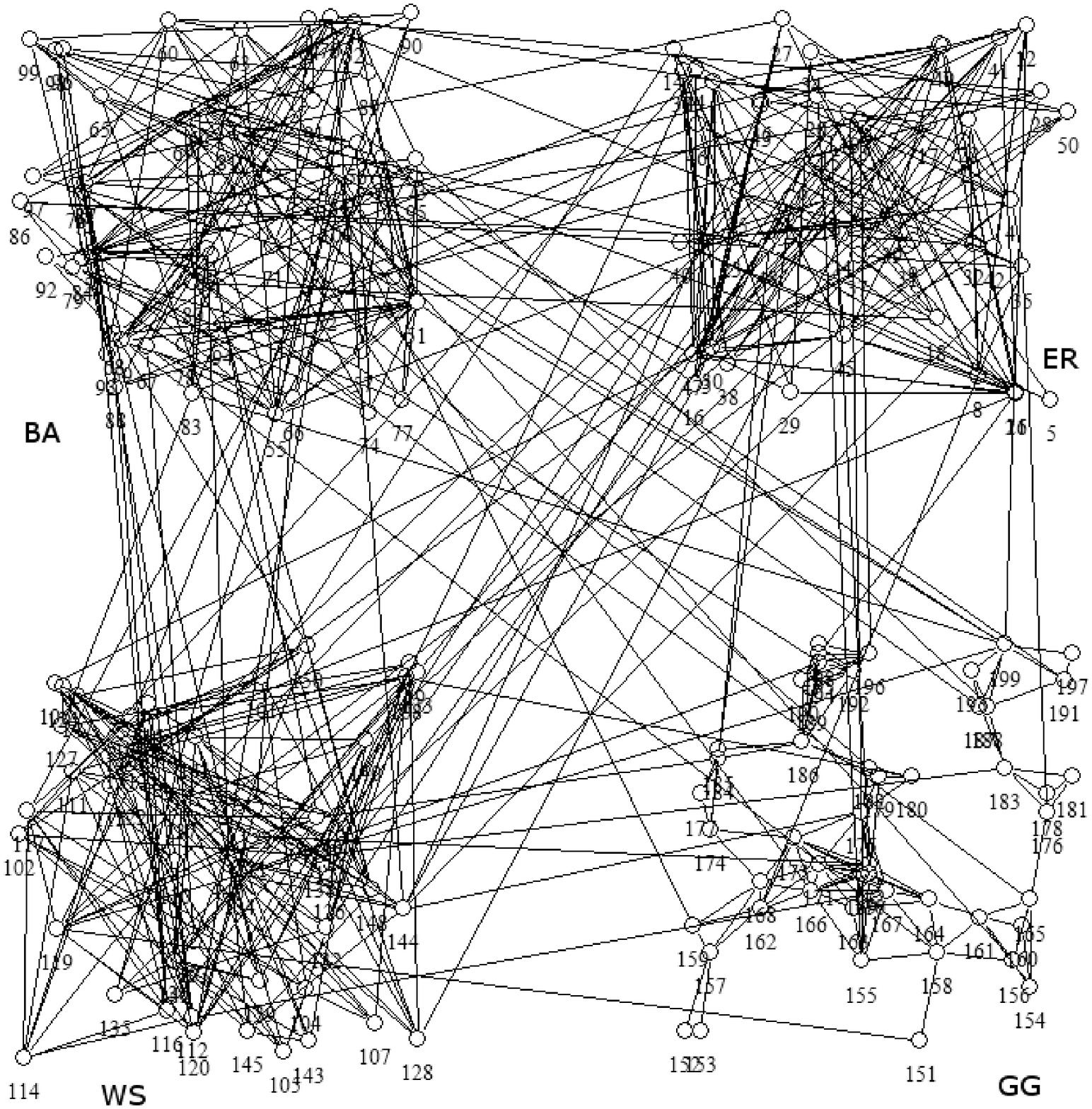} \\  
   (a) \\ \vspace{0.5cm}
  \includegraphics[width=0.7\linewidth]{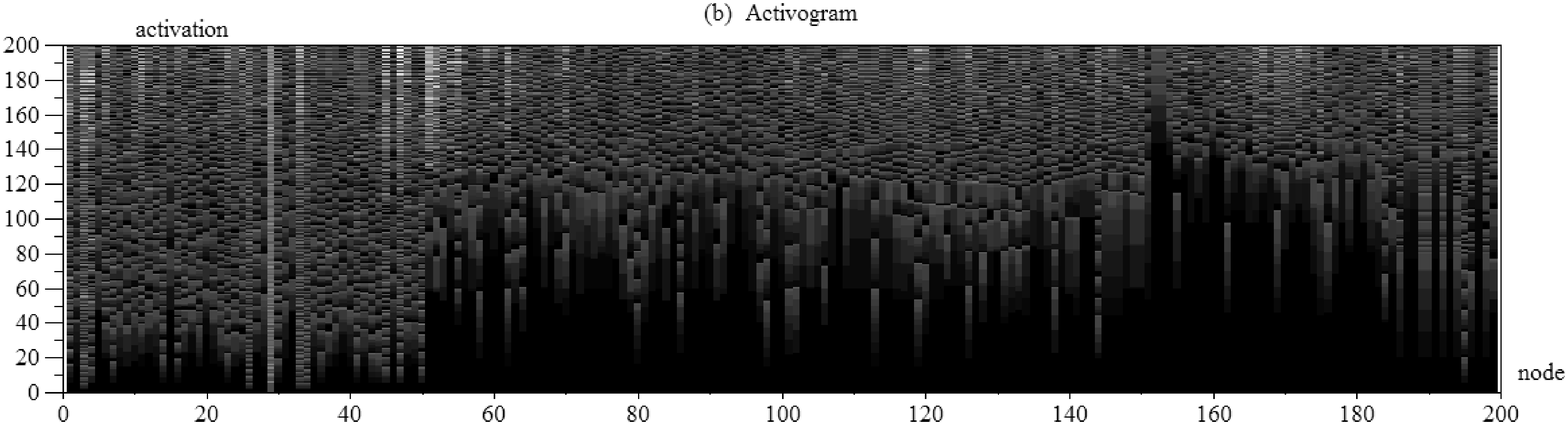} \\
  \vspace{0.5cm}
  \includegraphics[width=0.7\linewidth]{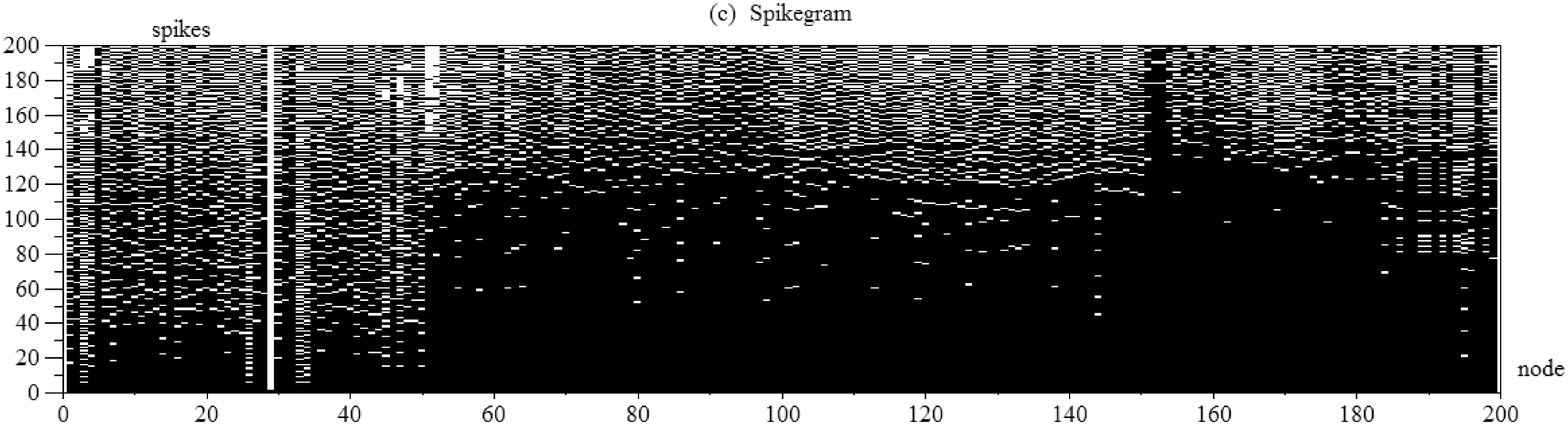} \\
  \vspace{0.5cm}
  \includegraphics[width=0.7\linewidth]{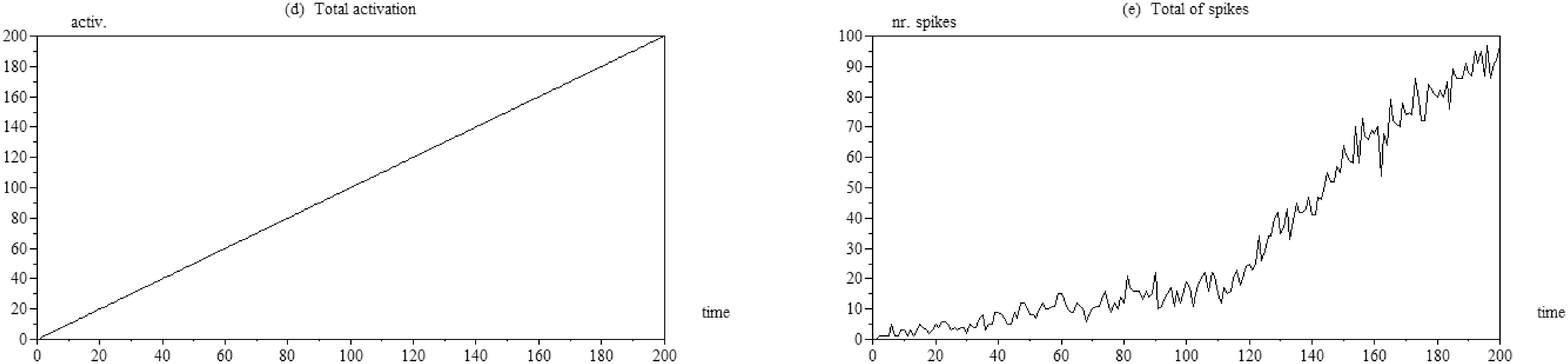} \\
  \vspace{0.5cm}
  \caption{A hybrid complex network containing four communities of
               different types (ER, BA, WS, and GG) is shown in
               (a).  The excitation of the network through one of
               its nodes $i$ tendd to activate first the community to which
               this node belongs to, confining the activation during a
               transient period of time.  For instance, the excitation
               of neuron $i=29$ induced an early activation of community
               1 in which it is included, as shown in the respective
               activogram (b) and spikegram (c). The total activation
               and spikes along time are shown in (d) and (e), respectively.
  }~\label{fig:ex_comm} 
  \end{center}
\end{figure*}

Interestingly, it was subsequently identified that the dissemination
of the activation emanating from the source node in integrate-and-fire
complex neuronal networks exhibiting modular structure tends to be
constrained inside the community to which the source node belongs.
This phenomenon is illustrated in Figure~\ref{fig:ex_comm} with
respect to a hybrid network (Fig~\ref{fig:ex_comm}a) composed of 4
communities of different type, namely Erd\H{o}s-R\'enyi (nodes 1 to
50), Barab\'asi-Albert (nodes 51 to 100), Watts-Strogatz (nodes 101 to
150) as well as a geographical structure (nodes 151 to 200).  Node 29,
which belongs to the Erd\H{o}s-R\'enyi community, was selected as the
source of activation, which was kept constant at intensity $1$.  The
respectively obtained activogram (namely the activation of each neuron
along the transient time) and spikegram (the spikes generated by each
neuron along time) are shown in Figures~\ref{fig:ex_comm}(b) and (c),
respectively.  The total activation and number of spikes along time
are shown in (d) and (e), respectively.

The activogram (Fig.~\ref{fig:ex_comm}b) makes it clear that community
1 (ER), extending from node 1 to 50, was activated first and remained
so along nearly 120 time steps, with the whole network undergoing an
avalanche of activation subsequently.  A similar behavior was obtained
for the spikes (Fig.~\ref{fig:ex_comm}c).  A few neurons from other
communities exhibited earlier activations and spiking.  These neurons
correspond to those implementing the inter-community connections.  As
expected, the total activation, shown in Figure~\ref{fig:ex_comm}(d),
increases linearly, reflecting the facts that activation is constantly
fed into the system through the source node and the conservation of
activation allowed by the adopted integrate-and-fire dynamics.  As
depicted in Figure~\ref{fig:ex_comm}, a completely different evolution
was obtained for the total number of spikes, which shows a clear surge
after nearly 120 time steps, corresponding to the main avalanche of
spikes taking place for most of the complex neuronal network.

Because of the close relationship between the communities and the
beginning activation times~\cite{Costa_begin:2008} and the average
activation~\cite{Costa_activ:2008} observed during the transient
regime, it was possible to use the patterns of spiking along the first
time steps in order to identify with accuracy the original
communities~\cite{Costa_begin:2008,Costa_activ:2008}.  More
specifically, the activation is placed at each node and the pattern
of beginning activation times or average activation are obtained and
then decorrelated by using the Principal Component Analysis
(e.g.~\cite{Costa_book:2001,Costa_surv:2007}) methodology.
Well-defined clusters are obtained in the projected space which
corresponds to the original communities, while the points between the
clusters corresponds to the edges which implement the intercommunities
connections.  Such a result has important implications for both
neuroscience and complex networks research.  In the former case, the
identified relationship between communities and integrate-and-fire
transient dynamics suggests that the integrate-and-fire dynamics of
biological neuronal cells is critical for confining the neuronal
activation inside the spatially clustered functional processing
modules of the brain (e.g.~\cite{Zeki:1999,Hubel:2005}).  Actually,
this relationship may play a key role in defining the functional
modules during the growth of the neural system as a consequence of
correlations between the presented stimuli.  The implications for
complex networks research are also many.  For instance, the method
seems to produce well-defined clusters even for relatively low
modularity indices, suggesting that it may be an effective
computational choice for community detection.  In addition, the
relationship between community structure and integrate-and-fire
dynamics lies at the heart of the important current investigations of
the relationship between complex networks structure and dynamics
(e.g.~\cite{Sznajd_BA, Newman:2003, Costa_vor:2004,
Fortunato_opinion, Boccaletti:2005, Zhou:2006,
Arenas_Vicente:2006, Lodato:2007, Ott:2007, Almendral:2007,
Ising_BA,COsta_Ising:2007}).  Indeed, the confinement phenomenon has
been observed (~\cite{Costa_begin:2008, Costa_activ:2008}) to depend
on the type of network topology, with geographical structures being
particularly poor in constraining activation.  For all such reasons,
it is important to investigate in more detail the origin and
properties of this interesting phenomenon.  This is precisely the main
objective of the current work.

It has been conjectured~\cite{Costa_begin:2008, Costa_activ:2008}
that the confinement of the activation/spiking in integrate-and-fire
complex neuronal networks is related to the avalanches of activations
which take place during the transient regime in the greatest part of
the analysed networks.  This seems to be supported by the fact that
the geographical structures, which do not produce avalanches, are less
likely to constrain the activation.  In the present work, we extend
the methodology for relating the avalanches to the hierarchical
organization of complex networks in order to address the activation
confinement inside communities.  The equivalent model described
in~\cite{Costa_equiv:2008} is enhanced to include the intra-level
activations and then extended to represent modular complex networks.
This has been accomplished by obtaining the hierarchical structure for
the network as a whole and then separating it into modules (see
also~\cite{Costa_dyn:2005}), on which the modular equivalent model was
based) corresponding to each original community.  By using a
combination of theoretical and experimental approaches involving the
above concepts, it has been possible to obtain decisive insights and
proper understanding of the origin and properties of the activation
confinement phenomenon in integrate-and-fire networks.

The article starts by reviewing the basic concepts and follows by
presenting the enhanced and extended equivalent model.  By considering
several hybrid networks, several key effects underlying the confirment
phenomenon are identified and discussed.  The article concludes by
identifying its main contributions and prospects for future developments.

\section{Basic Concepts}

\subsection{Complex Network Representation and Measurements}

In this work we consider undirected (the theoretical models used to
obtain the hybrid networks), directed (the respective extension to
complex neuronal networks), as well as unweighted and weighted (the
equivalent models) networks.  Weighted and directed networks are the
more general case from which the above specific structures can be
obtained~\cite{Costa_surv:2007}.  Such a type of network can be fully
represented in terms of its \emph{weight matrix} $W$, so that each
edge extending from node $i$ to node $j$ with respective weight $w$
implies $W(j,i)=w$.  The absence of edges from node $i$ to node $j$ is
expressed as $W(j,i)=0$.  The topology of a complex network described
in terms of its weight matrix can be immediately obtained by setting
to one all non-zero weights, which yields the \emph{adjacency matrix}.
Undirected networks can be obtained by symmetrizing all the directed
connections, so that each undirected edge incorporates two directed
edges with opposite directions.

The \emph{immediate neighbors} of a node $i$ are those nodes which
receive an edge from it.  The \emph{outdegree} of a node is the number
of its immediate neighbors.  The \emph{indegree} of a node $i$ is the
number of nodes which send edges to $i$.  If the network is
undirected, the indegree and outdegree are identical and called simply
\emph{degree}.  Two nodes are \emph{adjacent} if they share one edge. 
Two edges are adjacent if they share one node.  A sequence of adjacent
edges define a \emph{walk}.  A walk without repetition of nodes or
edges is a \emph{path}.  The \emph{length} of a walk or path is equal
to the number of edges they contain.

\subsection{Hierarchical Organization of Undirected and Directed 
Networks} \label{sec:hier}

Given a undirected network and one of its nodes (or even a
subgraph~\cite{Costa_EPJB:2005,Costa_dyn:2005}) as a reference, it is
possible to obtain its respective \emph{hierarchical organization with
respect to the chosen reference}.  Such an organization is defined by
the \emph{hierarchical} (or concentric) neighborhoods of the reference
node.  Let us call the reference node $i$.  Its first hierarchical
neighborhood includes the nodes which are directly attached to $i$,
i.e. the nodes which are adjacent to it.  The second hierarchical
neighbors are those which are at shortest path length 2 from $i$, and
so on.  The hierarchical structure of a network can be obtained by
flooding the network from the reference node~\cite{Costa:2004}.  More
specifically, the reference is labeled with value 1.  Then, the nodes
which are adjacent to the reference are labeled 2, and so on.

Given the hierarchical organization of a network with respect to one
of its nodes, it is possible to obtain a series of hierarchical
measurements (e.g.~\cite{Costa:2004, Costa_NJP:2007, Costa_JSP:2006,
Costa_EPJB:2005}) which provide information not only about the
immediate neighborhood of a node, but about the whole set of
topological scales from the immediate node level to the complete
network level.  The properties of the hierarchical organization of
several theoretical and real-world networks have been systematically
investigated in previous works (e.g.~\cite{Costa:2004, Costa_NJP:2007,
Costa_JSP:2006, Costa_EPJB:2005}).  The generalization of the
hierarchical approach to consider subgraph references has been
described in~\cite{Costa_EPJB:2005}.

Because in this work we will be dealing with directed networks, it
is important to consider the generalization of the hierarchical
concepts to this type of networks.  This can be immediately achieved
by considering the flooding algorithm outlined above.  Given the
reference node, the hierarchies along the directed network is
identically obtained by the same algorithm.  However, while the
outgoing connections in the obtained hierarchical structure are always
found between successive concentric levels (i.e. extending from level
$h$ to level $h+1$, with necessarily no connections between $h$ and
any other concentric level), the backward connections can take place
between any concentric levels (e.g. it is possible to have directed
edges extending from level $h$ to any of the previous levels).

\subsection{Hybrid Modular Networks} \label{sec:hybr}

In this work we will consider 4 theoretical complex networks models
(e.g.~\cite{Albert_Barab:2002, Dorogov_Mendes:2002, Newman:2003,
Boccaletti:2006, Costa_surv:2007}) as corresponding to communities:
Erd\H{o}-R\'enyi (ER, see also~\cite{Flory}), Barab\'asi-Albert (BA),
Watts-Strogatz (WS), as well as a simple geographical model (GG).  The
ER structures are obtained by connecting pair of edges with constant
probability.  A BA network is obtained~\cite{Albert_Barab:2002} by
starting with $m0$ nodes and incorporating new nodes with $m$ edges
each, which are attached to the previous nodes with probability
proportional to their respective degrees. The WS structures considered
in this work starts as linear regular lattices of suitable degree,
followed by rewiring of $10\%$ of the edges.  The GG networks are
obtained by distributing nodes along a two-dimensional space and
connecting all nodes which are closer than a maximum distance.  In all
cases, only the largest connected component was used.  Similar
intra-community degrees are adopted henceforth.

Informally speaking, a \emph{community} in a network is one of its
subgraphs characterized by the fact that its nodes are more intensely
connected one another than with the remainder of the network.  A
complex network incorporating communities is henceforth called a
\emph{modular network}.  A \emph{hybrid modular network} involves 
communities of different topological types.  In addition to specifying
the type of each community, it is also necessary to specify the inter
and intra-community degrees.  Figure~\ref{fig:ex_comm}(a) illustrates
a hybrid network composed of 4 communities of respective ER, BA, WS
and GG types.  Each of these communities have 50 nodes.  The
inter-community degree is 1 (i.e. each two nodes are connected, in
the average, with one node from another community) and intra-community
degree equal to 6 (i.e. each node connects to 6 other nodes, in the
average, inside its respective community).  Hybrid modular networks
are particularly useful for community studies because they allow the
identification of the effects of their respective topologies on the
community structure and separation.

\subsection{Integrate-and-Fire Complex Neuronal Networks}

\begin{figure}[htb]
  \vspace{0.3cm} 
  \begin{center}
  \includegraphics[width=1\linewidth]{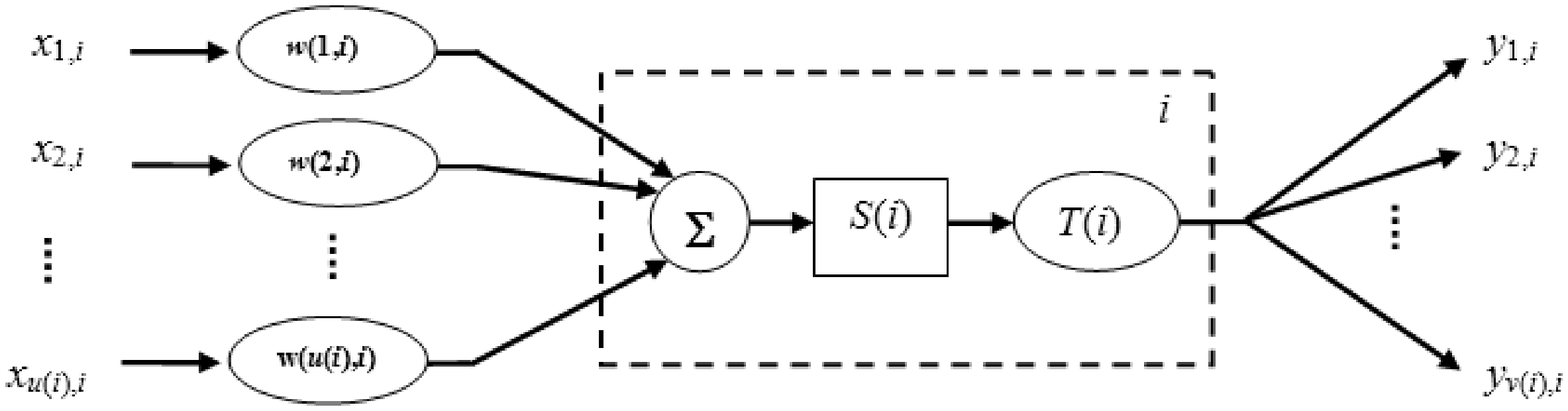} \\
  \vspace{0.5cm}
  \caption{The integrate-and-fire model of neuronal cell adopted
                in this work.
  }~\label{fig:neuron} 
  \end{center}
\end{figure}

Figure~\ref{fig:neuron} illustrates the basic integrate-and-fire model
of neuron adopted in this work.  It includes a set of weights, an
integrator, a memory $S(i)$ containing the internal activation, and a
non-linear transfer function, namely a hard-limit function with
threshold $T(i)$.

Because the original hybrid modular networks obtained as described in
Section~\ref{sec:hybr} are undirected, it is necessary to transform
them into respective directed versions in order to be used as complex
neuronal networks.  This can be immediately achieved by splitting each
undirected edge into two directed edges, corresponding to a dendrite
and an axon.  Observe that all the nodes in the resulting directed
complex neuronal networks consequently have indegree equal to the
outdegree.  All hybrid complex networks considered in this article
have weights equal to 1. The equivalent models, however, use varying
weights in order to implement the proportional splitting of activation
between hierarchical levels.  Also, while all thresholds in the hybrid
complex neuronal networks are identical to 1, the thresholds in the
equivalent models are set so as to reproduce the inertia of the number
of nodes associated to each equivalent node.

The source of activation is placed at a specific node of the network
(it is also possible to consider activations emanating from groups of
nodes).  The activation fed by the source into the complex neuronal
networks are constant and equal to 1, implying linear increase of the
total activation along time (see Figure~\ref{fig:ex_comm}d).  All
activations arriving at a given neuron are integrated (remember all
weights are equal to 1) along time (facilitation~\cite{Squire:2003}),
being stored into the internal memory $S(i)$.  Once this integrated
value exceeds the threshod, i.e. $S(i)\geq T(i)$, the neuron fires,
liberating the stored activation, which is equally distributed among
the axons.  Such an operation ensures conservation of the total
activation.  Though the liberation of the total activation is not
biologically realistic (but can be emulated by setting adequate
thresholds), similar phenomena of avalanches and activation
confinement inside communities have been identified for complex
neuronal netwrosk where the outgoing activation is limited to the
threshold value, with the difference that the avalanches express
themselves as peaks of spiking, instead of abrupt increases as in
Figure~\ref{fig:ex_comm}(e).

\section{The First Equivalent Model}  \label{sec:first}

The equivalent model of activation of integrate-and-fire complex
neuronal networks described in~\cite{Costa_equiv:2008} was allowed
because the activation of nodes in such systems is strongly determined
by the hierarchical organization of the respective structures.  After
choosing the source node $i$, this model is obtained as follows.  By
using the methodology described in~\cite{Costa:2004, Costa_NJP:2007,
Costa_JSP:2006, Costa_EPJB:2005} we calculate: the hierarchical number
of nodes $n_h(i)$ (i.e. the total number of connections from the
neurons in level $h$ to the neurons in level $h+1$) as well as the
hierarchical degree $k_h(i)$ (i.e. the number of nodes within the
concentric level $h$).  The equivalent nodes are defined as subsuming
the set of nodes at each respective concentric level of the original
network (see also~\cite{Costa_dyn:2005}).  Because the networks
originally considered had symmetric connections (the undirected edges
were symmetrized by splitting them into two directed edges), the only
possible connections between the equivalent nodes take place between
adjacent concentric levels, i.e. the node at level $h$ connects only
to nodes $h-1$ and $h+1$.  Such a connectivity implies the equivalent
model to be a chain of the equivalent nodes.  The weights of the
connections between adjacent nodes are defined as

\begin{eqnarray}
  W(h+1,h) = k_h(i)/d \nonumber \\
  W(h-1,h) = k_{h-1}(i)/d  \nonumber \\
  \nonumber
\end{eqnarray}

where $d = k_h(i) +k_h{h-1}$ for proper normalization.  Thus, the
weights represent the fraction of the activation which is sent to the
left and right neighbors of the equivalent nodes in the chain model.
Such a model is then activated from the source node $i$.

Assuming a reasonable level of degree regularity in the network, the
neurons at each hierarchical level will tend to fire almost
simultaneously after a period of activation integration which is
proportional to the sum of the number of nodes from the hierarchical
level to the current level.  So, the \emph{beginning activation times}
of the neurons at each concentric levels can be predicted from such
sums of nodes.  The main avalanche tends to take place when the
neurons in the concentric level with the largest number of nodes
spike.  Because the activation liberated by this event is very large,
it tends to fire all neurons along the adjacent layers (in both
directions) within a brief period of time.  The intensity of the main
avalanche can be estimated from the number of neurons at the level
with the largest number of nodes and at its adjacent levels.  The time
at which the avalanche takes place can be predicted by adding the
number of neurons from the concentric level 1 up to the critical
level.

\section{Enhancing the Equivalent Model}

In this section we describe how to enhance the chain equivalent model,
previously suggested in~\cite{Costa_equiv:2008}, in order to
incorporate the intra-ring edges within each concentric level.  This
can be immediately done by incorporating self-connections in the chain
equivalent model, in addition to the weights respective to the
connections from level $h$ to levels $h+1$ and $h-1$, given by the
hierarchical degrees $k_h(h)$ and $k_h{h-1}$.  In order to properly
normalize the distribution of activation from each level $h$, the
respective weights are now redefined as

\begin{eqnarray}
  W(h+1,h) = k_h(i)/d \nonumber \\
  W(h-1,h) = k_{h-1}(i)/d  \nonumber \\
  W(h,h) =   a_h(i)/d   \nonumber  \\ \nonumber
\end{eqnarray}

where $d = k_h(i) +k_h{h-1}+a_h(i)$, $k_h(i)$ is the hierarchical
degree of level $h$ with respect to reference node $i$ and $a_h(i)$ is
the intra-ring degree of level $h$ with respect to node $i$.  The
latter measurement expresses the number of edges within the concentric
level $h$.

The incorporation of the self-connections into each concentric level
is expected to provide a more complete model of the dynamics unfolding
along the hierarchical organization of the original network.  More
specifically, as hinted in~\cite{Costa_equiv:2008}, the intra-ring
edges tend to reduce the distribution of its current activation to the
other levels by deviating a portion $W(h,h)$ of activation towards the
same level.  By doing so, the intra-ring connections tend to delay the
nearly-simultaneous spiking of the respectively concentric levels.  In
case the intra-ring edges are not distributed uniformly between the
nodes of a given concentric level, the neurons of the next layer will
tend to spike in a less simultaneous way, implying a smoothing of the
number of spikings and avalanches.

\begin{figure*}[htb]
  \vspace{0.3cm} 
  \begin{center}
  \includegraphics[width=0.35\linewidth]{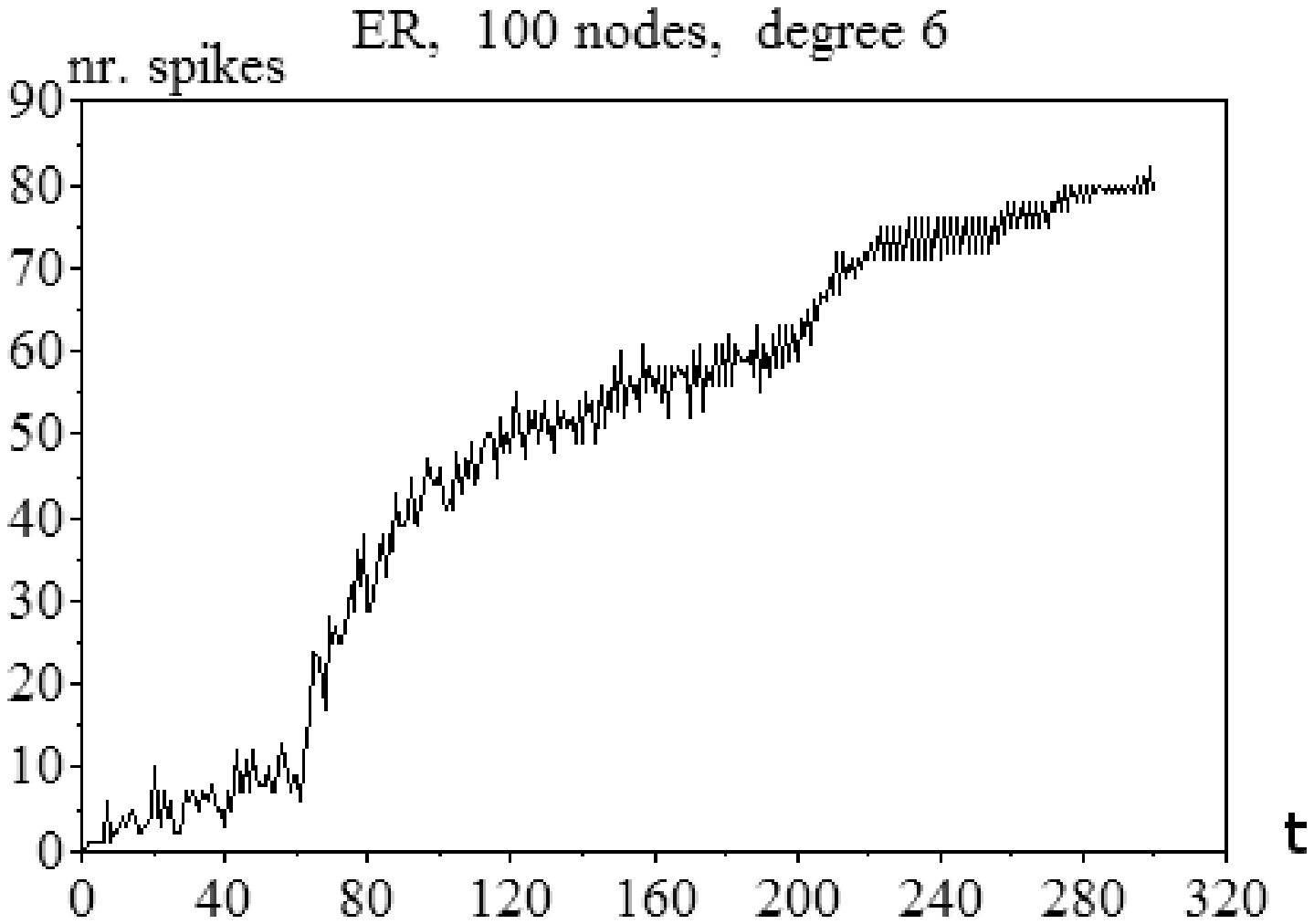}   
  \includegraphics[width=0.35\linewidth]{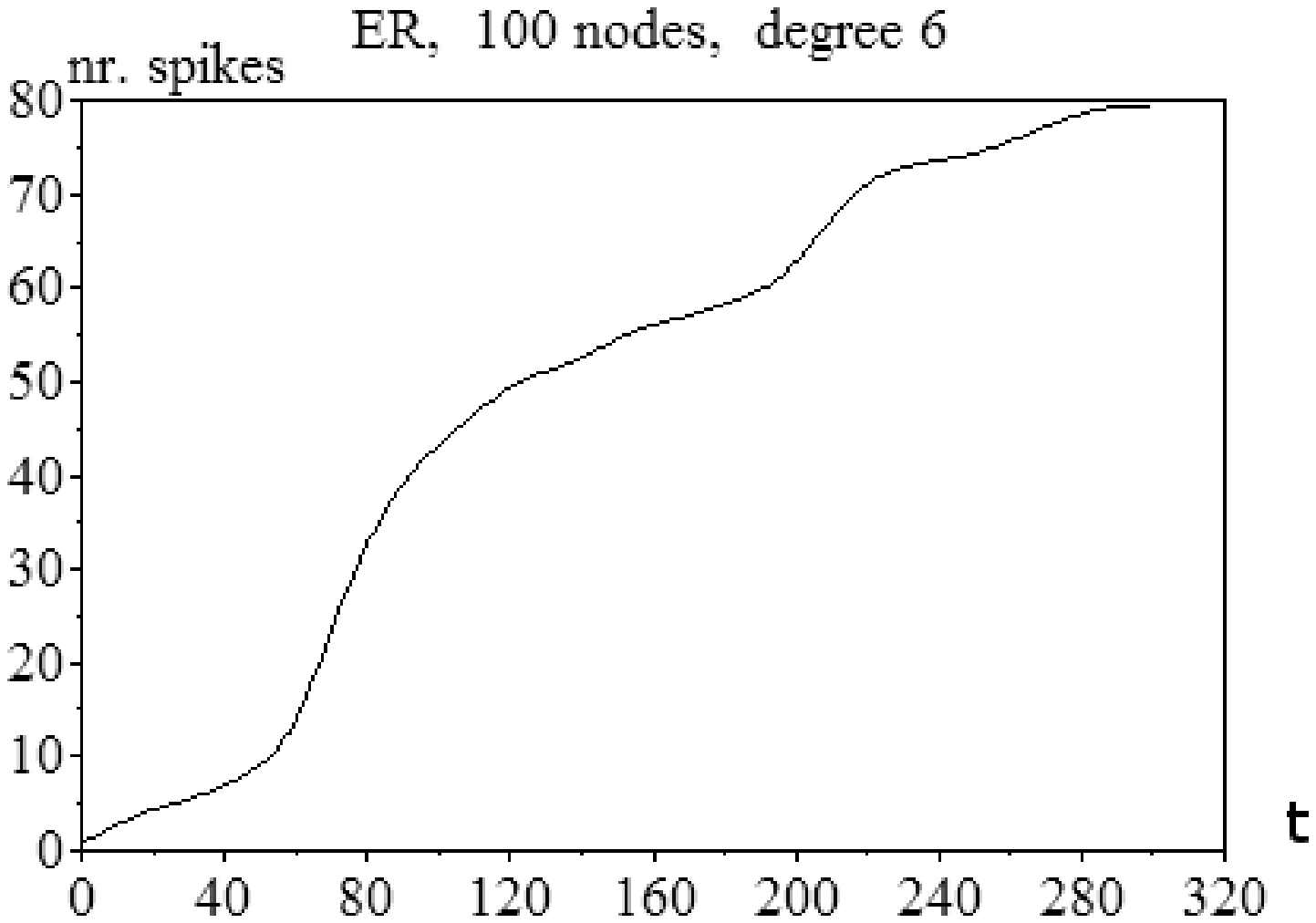}   \\
   (a)  \hspace{5.5cm} (b) \\ \vspace{0.5cm}
  \includegraphics[width=0.35\linewidth]{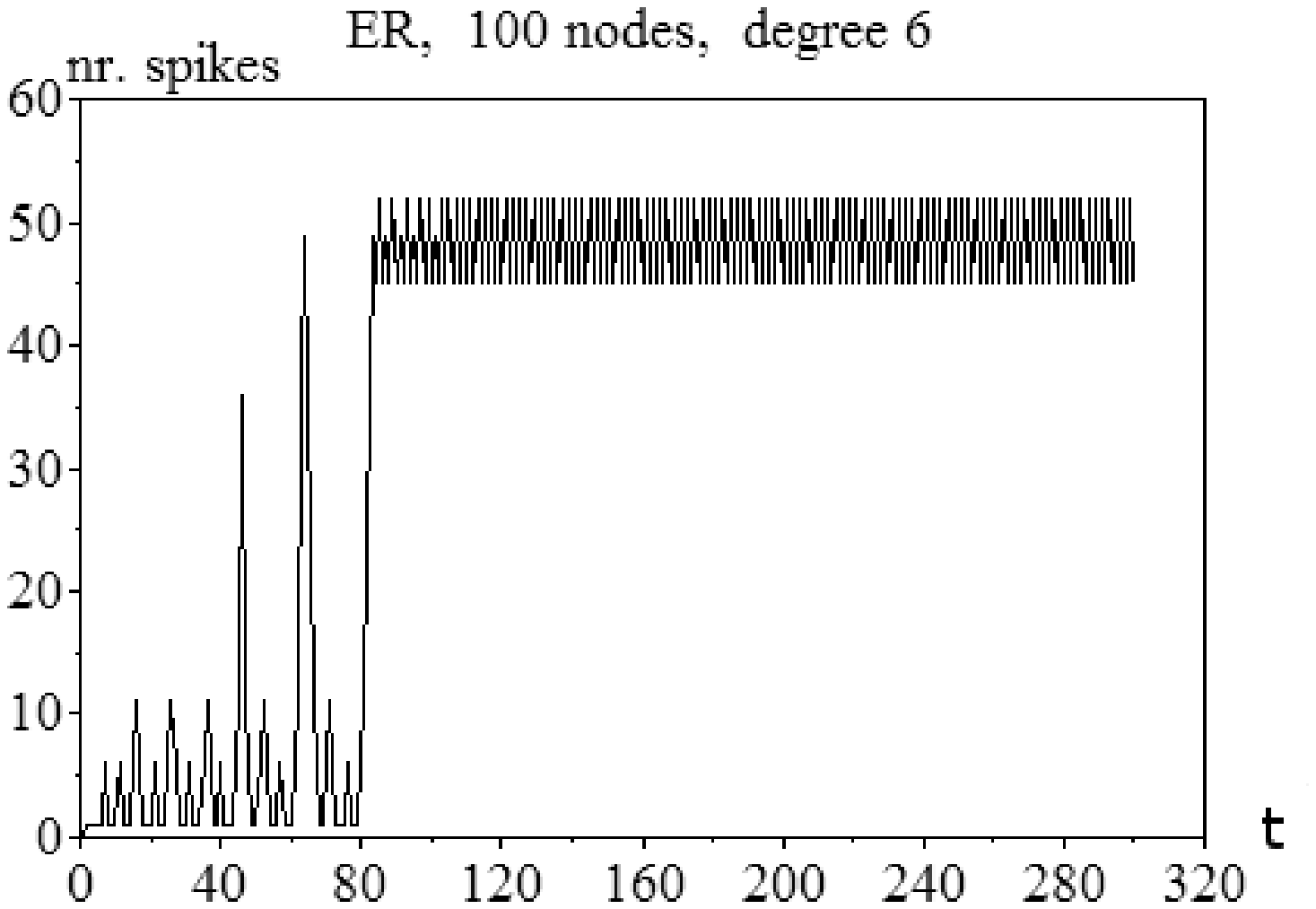}   
  \includegraphics[width=0.35\linewidth]{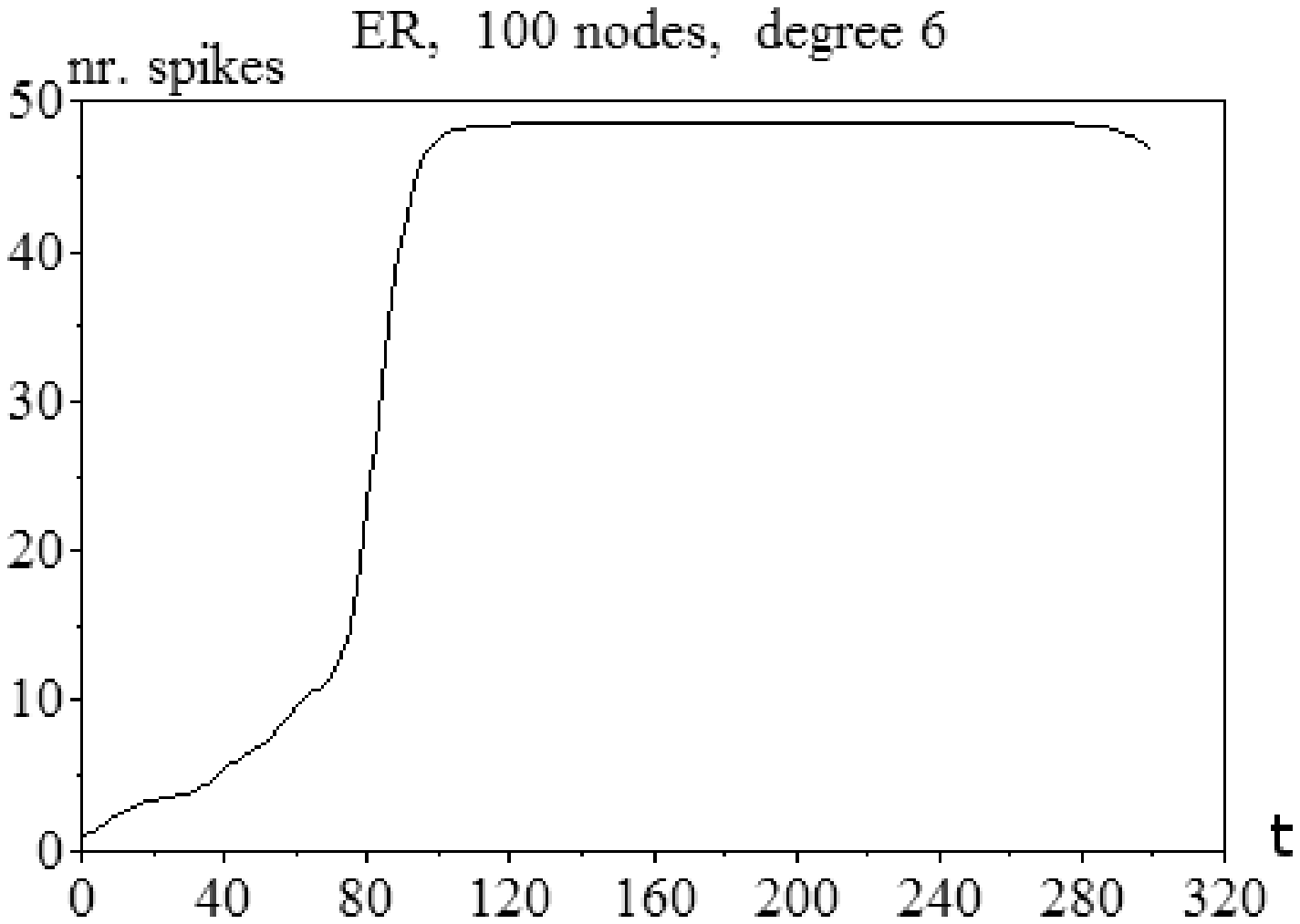}   \\
  (c)  \hspace{5.5cm}  (d)  \\
  \includegraphics[width=0.35\linewidth]{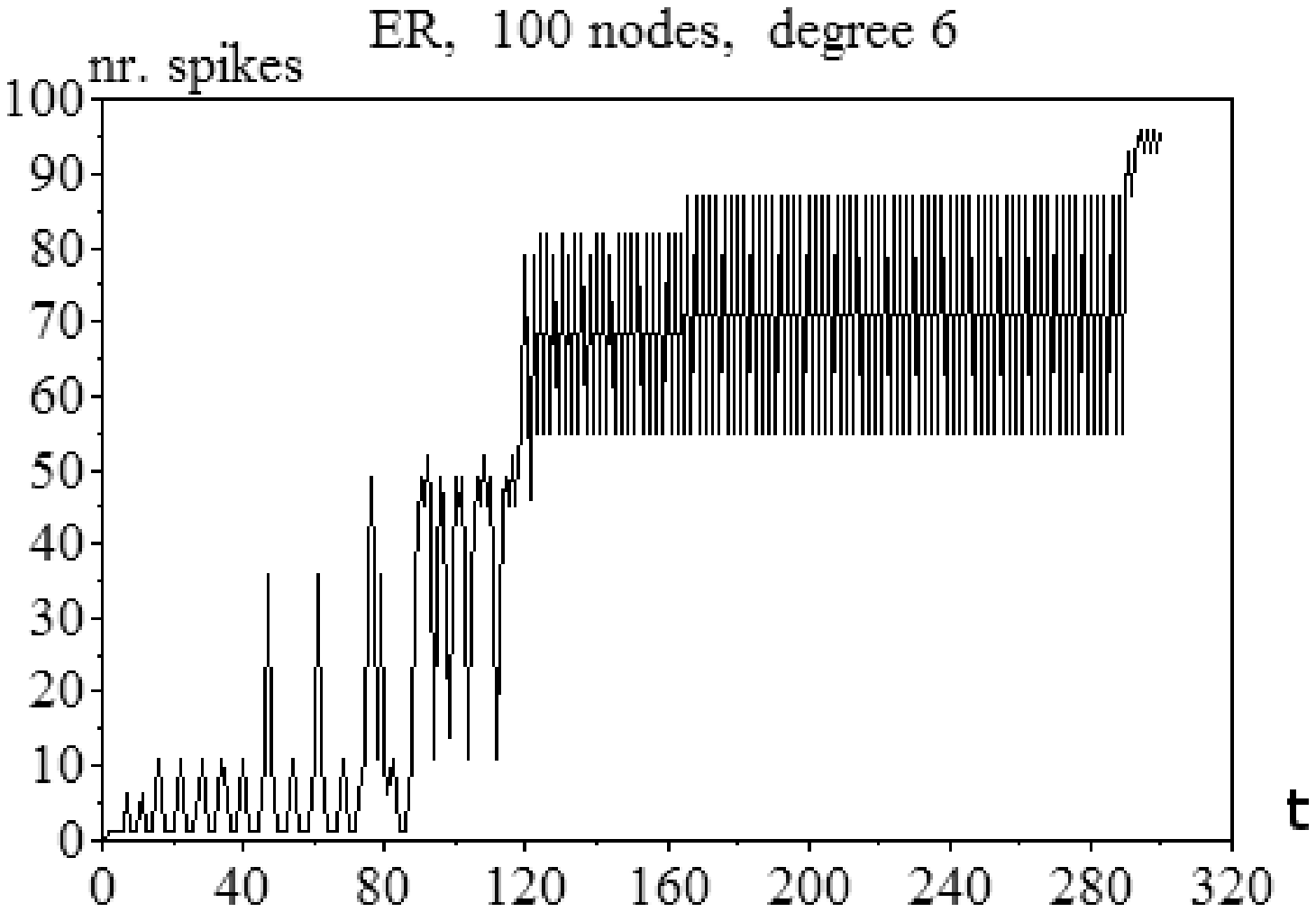}   
  \includegraphics[width=0.35\linewidth]{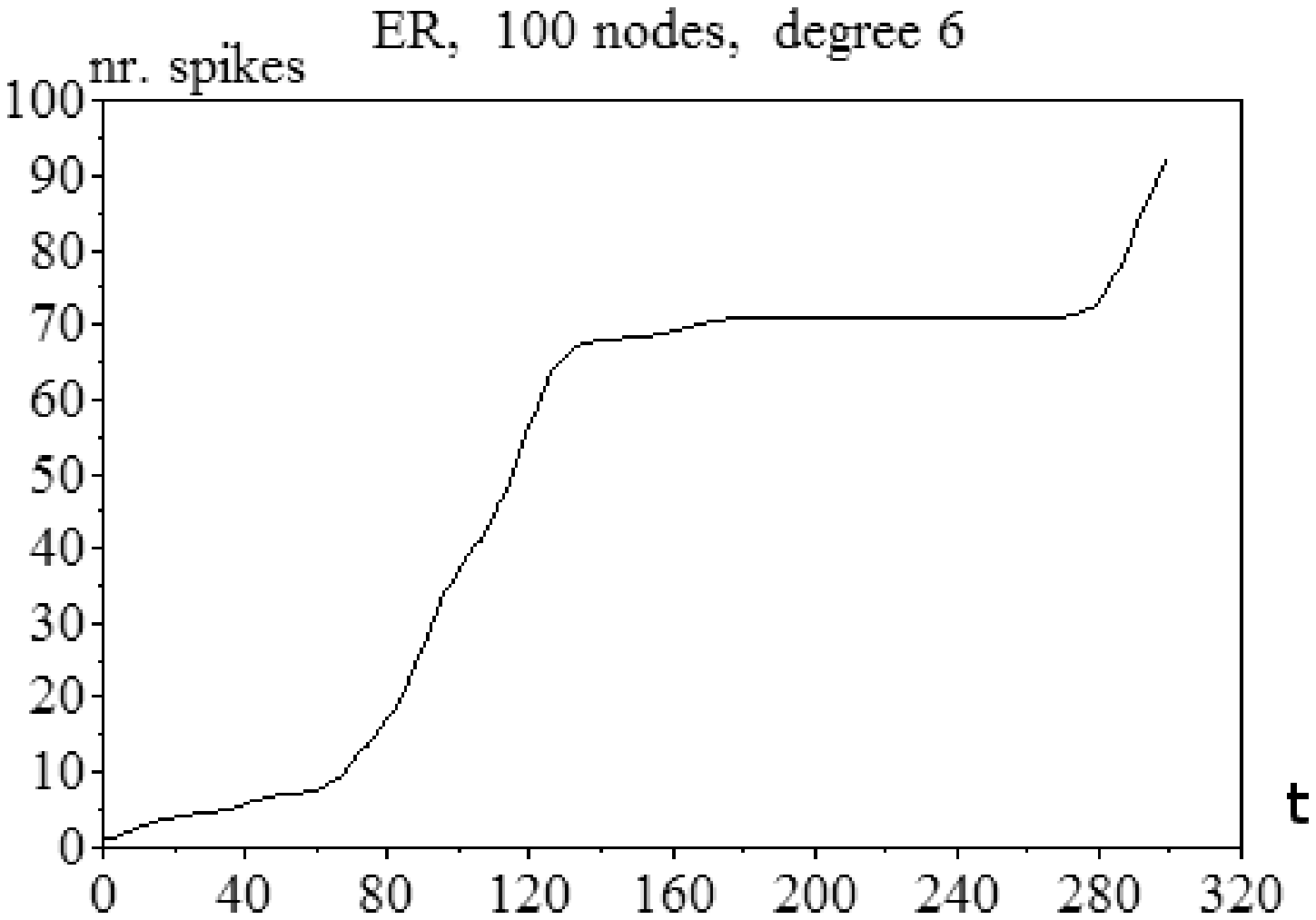}   \\
  (e)  \hspace{5.5cm}  (f)  \\
  \vspace{0.5cm}
  \caption{The total number of spikes along time obtained for a
             complete original ER network with 100 nodes and
             average degree equal to 6 (a), and its respectively
             smoothed version (b).  The total number of spikes as
             estimated by the chain equivalent model while not
             taking into account the intra-ring connections (c),
             and its respectively smoothed version.  The consideration 
             of the intra-ring connections enhanced the estimation
             of the number of spikes as shown in (e) and the
             respectively smoothed version (f).  The smoothings
             were performed by convolving the original curves with
             a Gaussian function with standard deviation equal to
             8 time steps.
  }~\label{fig:ex_intra} 
  \end{center}
\end{figure*}

Figure~\ref{fig:ex_intra} illustrates the effect of incorporating the
intra-ring degrees while modeling a traditional (i.e. without
communities) ER network with 100 nodes and average degree equal to 8.
Figure~\ref{fig:ex_intra}(a) shows the total number of spikes obtained
for the whole original network, while Figure~\ref{fig:ex_intra}(b)
shows the respectively smoothed version.  Because the curves showing
the number of spikes tend to be jagged as a consequence of the saw
oscillations discussed in~\cite{Costa_equiv:2008}, it is interesting
to smooth the activation/spiking signals.
Figure~\ref{fig:ex_intra}(b) was obtained by convolving the curve in
Figure~\ref{fig:ex_intra}(a) with a Gaussian function with standard
deviation equal to 6 time steps.  The total number of spikes produced
by the respective chain equivalent model without consideration of the
intra-ring degrees (as adopted in~\cite{Costa_equiv:2008}) is shown in
Figure~\ref{fig:ex_intra}(c), with its respective smooth version
depicted in (d).  As expected, the elimination of the intra-ring
connections tends to enhance the simultaneity of the spikings,
producing a more definite and abrupt avalanche.  The results obtained
by the enhanced chain equivalent model incorporating self-connections
are shown in Figures~\ref{fig:ex_intra}(e) and (f) respectively to the
original signal and its smoothed version.  It is clear that the
incorporation of the intra-ring degrees as the weights of the
self-connections in the chain equivalent model led to a substantially
more precise estimation of the original spiking dynamics
(Fig.~\ref{fig:ex_intra}a and b).

The effect of the intra-ring connections in reducing the simultaneity
of the spiking at each concentric level and dispersing the avalanche
(i.e. implying a smoother transition along time) suggests that it
would be possible to enhance the simultaneous activation of the
communities while the network is fed from the source node, possibly
enhancing further the ability of the method to detect communities.
Such a possibility is explored further in Section~\ref{sec:mod}.

\section{Generalizing the Equivalent Model to Non-Symmetric Connectivity}

\begin{figure}[htb]
  \vspace{0.3cm} 
  \begin{center}
  \includegraphics[width=0.9\linewidth]{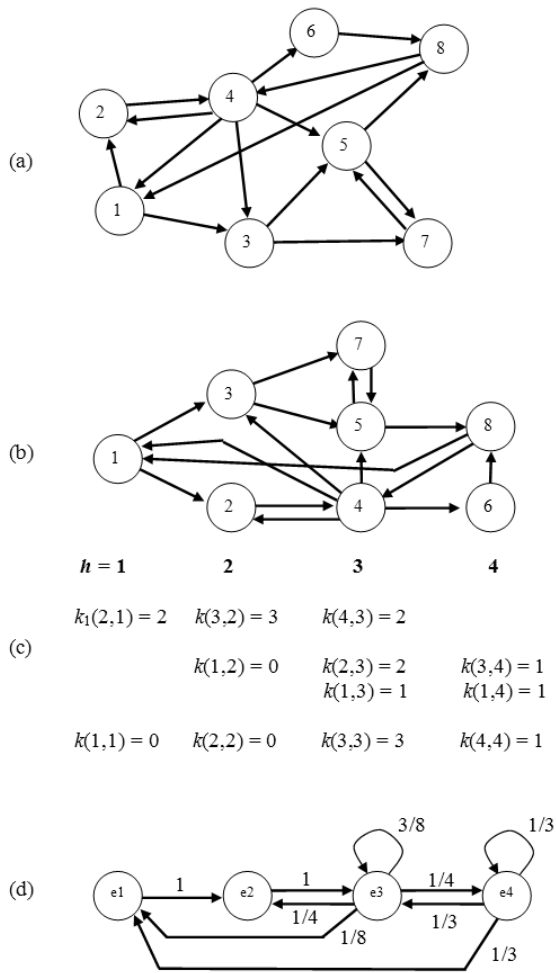} \\
  \vspace{0.5cm}
  \caption{Example of how to obtain the generalized equivalent
             model from an asymmetrically directed complex
             neuronal network.  The weights are obtained from 
             the generalized hierarchical degrees.  For instance,
             the weight of the connection from equivalent node
             $e_4$ to the equivalent node $e_1$ is obtained by
             dividing the number of edges departing from $e_3$
             towards $e_1$  (i.e. 1 edge) by the total number of 
             edges departing from $e_3$ towards any other equivalent 
             node except itself (i.e. 8 edges).
  }~\label{fig:gen} 
  \end{center}
\end{figure}

As mentioned in Section~\ref{sec:first}, the first equivalent model
assumed the edges of the original network to be symmetric in the sense
of one incoming directed edge matching every outgoing directed edge.
Such a connectivity was immediately implied by the splitting of the
originally undirected edges into two directed edges.  In this section
we generalize the equivalent model in order to allow the consideration
of asymmetric directed connections, i.e. any directed complex neuronal
network, by considering the hierarchical organization of directed
complex networks described in Section~\ref{sec:hier}.  Such a
generalization is explained with the help of Figure~\ref{fig:gen},
which also illustrates the previously introduced concepts of
self-connections.

Figure~\ref{fig:gen}(a) shows a simple directed network involving
several asymmetric connections.  Let us consider node 1 as the
reference node.  By flooding the network from this node, the
hierarchical layers are defined by the nodes which are simultenously
flooded at each time, yielding the hierarchical organization shown in
Figure~\ref{fig:gen}(b).  Alternatively, the nodes at hierarchical
level $h$ are those which are at shortest directed path distance $h$
from the reference node $i=1$.  Observe that the forward directed
edges (i.e. those edges connecting from level $i$ onwards) are only
allowed between subsequent levels, i.e. the nodes in hierarchical
level $h$ can send forward connections only to the immediately
adjacent level $h+1$.  Contrariwise, the backward connections are
allowed from each level $h$ to any of the previous levels (e.g. the
connection from node 8 to node 1).  Observe also the existence of the
intra-ring connections at levels 3 and 4.  Because of such a more
generalized connectivity, the hierarchical degrees and intra-ring
degrees are redefined as the \emph{generalized hierarchical degrees}
$w(e_g,e_h)$, expressing the number of edges sent from the equivalent
node $e_h$ to the equivalent node $e_g$.  The generalized hierarchical
degrees of the directed network in Figure~\ref{fig:gen}(a) are shown
in Figure~\ref{fig:gen}(c).  Figure~\ref{fig:gen}(d) depicts the
respectively obtained equivalent model, incorporating the
self-connections.  The weights of the edges can be obtained as

\begin{equation}
  W(g,h) = k(e_{g},e_{h})/d  \nonumber \\ \nonumber
\end{equation}

where $d = \sum_{g \in \Omega(h)} k(e_{g},e_{h})$ and $\Omega(h)$ is
the set of equivalent nodes which receive a directed edge from
$e_h$. Obseve that $W(b,a)$ expresses the weight of the connection from
node $a$ to node $b$.

\section{The Modular Equivalent Model}

\begin{figure}[htb]
  \vspace{0.3cm} 
  \begin{center}
  \includegraphics[width=0.7\linewidth]{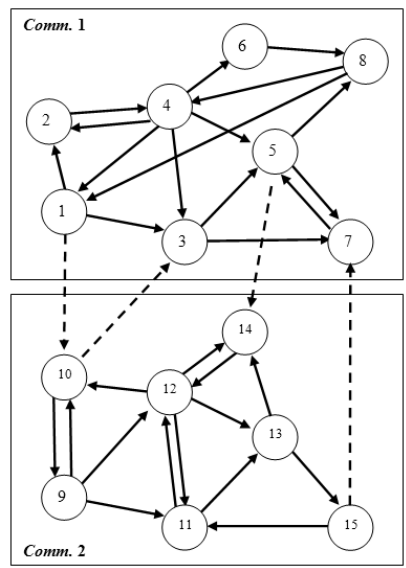} \\
  \vspace{0.5cm}
  \caption{A simple modular network composed of two communities.
  }~\label{fig:mod} 
  \end{center}
\end{figure}

\begin{figure}[htb]
  \vspace{0.3cm} 
  \begin{center}
  \includegraphics[width=0.7\linewidth]{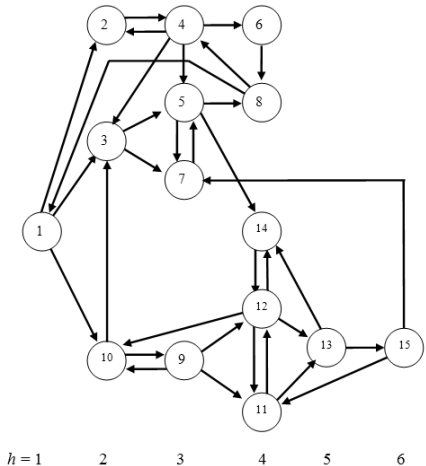} \\
  \vspace{0.5cm}
  \caption{The hierarchical organization of the module network
                in Figure~\ref{fig:mod} assuming node 1 as reference.
  }~\label{fig:merged} 
  \end{center}
\end{figure}

Having enhanced and generalized the original chain equivalent model,
we are now in a position to extend it to modular networks, i.e. complex
networks which incorporate communities.  This will be done with
respect to an example considering the simple modular network in
Figure~\ref{fig:mod}.  This network includes two respectively
identified communities, with 4 inter-community connections.  The first
step is to select the reference node so that the hierarchical
organization of the whole network can be determined as discussed
before.  We select node 1 as reference and source of activation.  The
hierarchical organization of the network in Figure~\ref{fig:mod} is
shown in Figure~\ref{fig:merged}.  A total of 6 concentric levels were
identified for this modular network.

Thus far, we have simply adopted the typical approach for the
hierarchical analysis of the network with respect to a given reference
node, without considering its community structure.  We do that now by
defining each equivalent node $e_{c,h}$ as subsuming the nodes of the
hierarchical level $h$ which belong to community $c$.  For instance,
in the case of the network in Figure~\ref{fig:mod}, we have that
$e_{1,3}$ is equal to 3, because there are 3 nodes of community $c=1$
at the concentric level $h=3$.  The generalized hierarchical degrees
of the equivalent nodes, henceforth expressed as $k(e_{d,g},e_{c,h})$,
are defined as the number of edges in the original network going from
the nodes corresponding to the equivalent node $e_{c,h}$ to $e_{d,g}$.
The weights of the connections between the equivalent nodes can now be
calculated as 

\begin{eqnarray} \label{eq:weights}
  W(e_{d,g},e_{c,h}) = k(e_{d,g},e_{c,h}) /d  \\  \nonumber
\end{eqnarray}

where $d = \sum_{g \in \Omega(h)} k(e_{d,g},e_{c,h})$ and $\Omega(h)$
is the set of equivalent nodes which receive a directed edge from
$e_{c,h}$.  The thresholds of each equivalent node are given by their
respective number of original nodes.  The specification of the
thresholds completes the definition of the modular equivalent model.

\begin{figure}[htb]
  \vspace{0.3cm} 
  \begin{center}
  \includegraphics[width=1\linewidth]{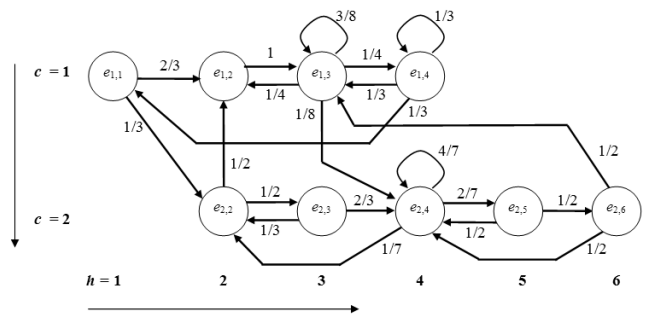} \\ \vspace{0.5cm}
  \caption{The modular equivalent model obtained for the network
                in Figure~\ref{fig:mod}.
  }~\label{fig:eq_mod} 
  \end{center}
\end{figure}

Figure~\ref{fig:eq_mod} shows the modular equivalent model obtained
for the network in Figure~\ref{fig:mod}.  Observe that the modular
structure (i.e. each original community) has been mapped into the rows
of this figure, while the hierarchical levels ($h=1, 2,
\ldots, 6$) succeed one another along the columns.  It is important to 
observe that this model corresponds to an integrate-and-fire complex
neuronal network by itself, but with weights not necessarily equal to
1.

\section{Activation Confinement in Hybrid Modular Networks} \label{sec:mod}

\begin{figure*}[htb]
  \vspace{0.3cm} 
  \begin{center}
  \includegraphics[width=0.4\linewidth]{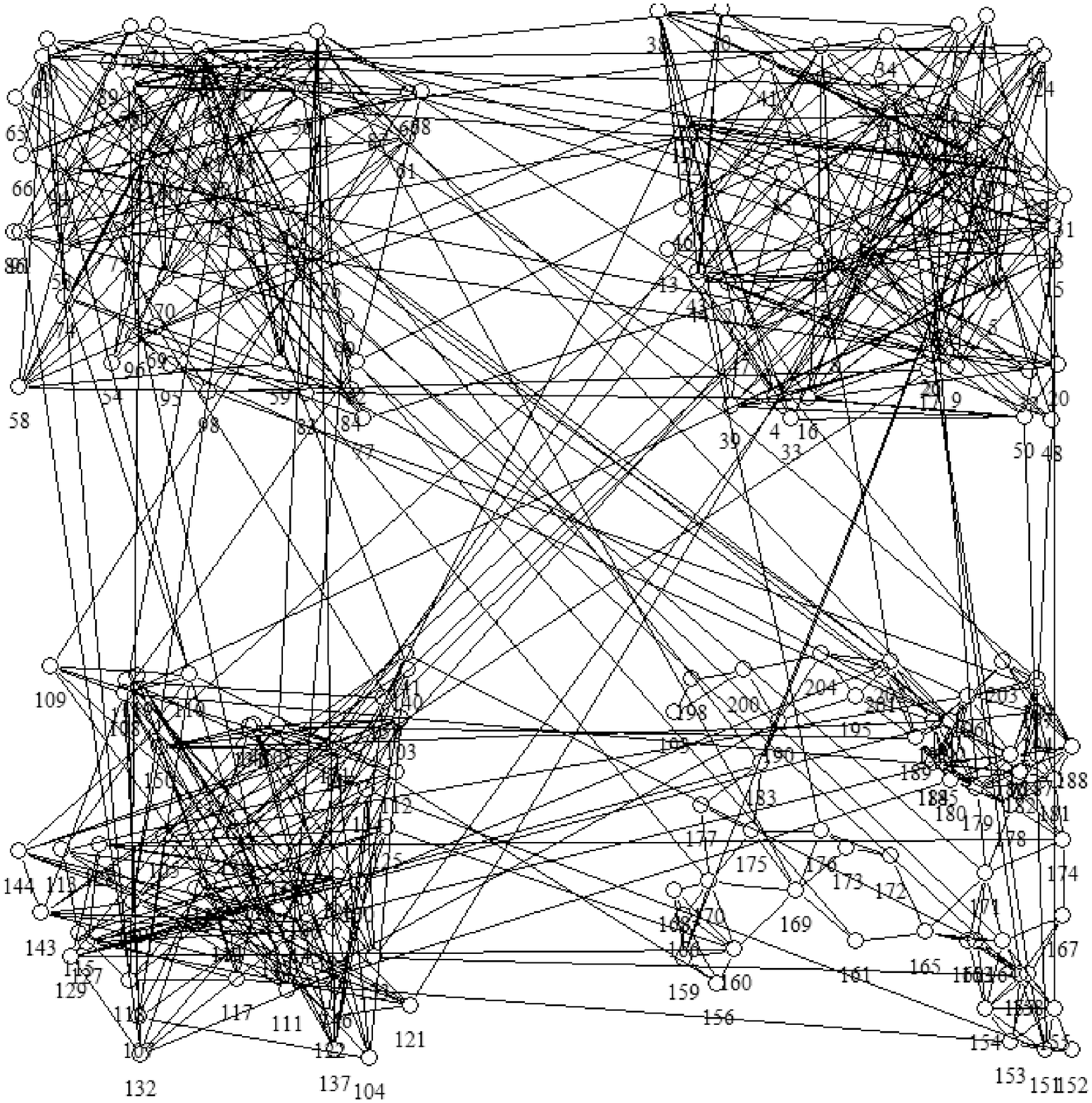}  \hspace{2cm}
  \includegraphics[width=0.4\linewidth]{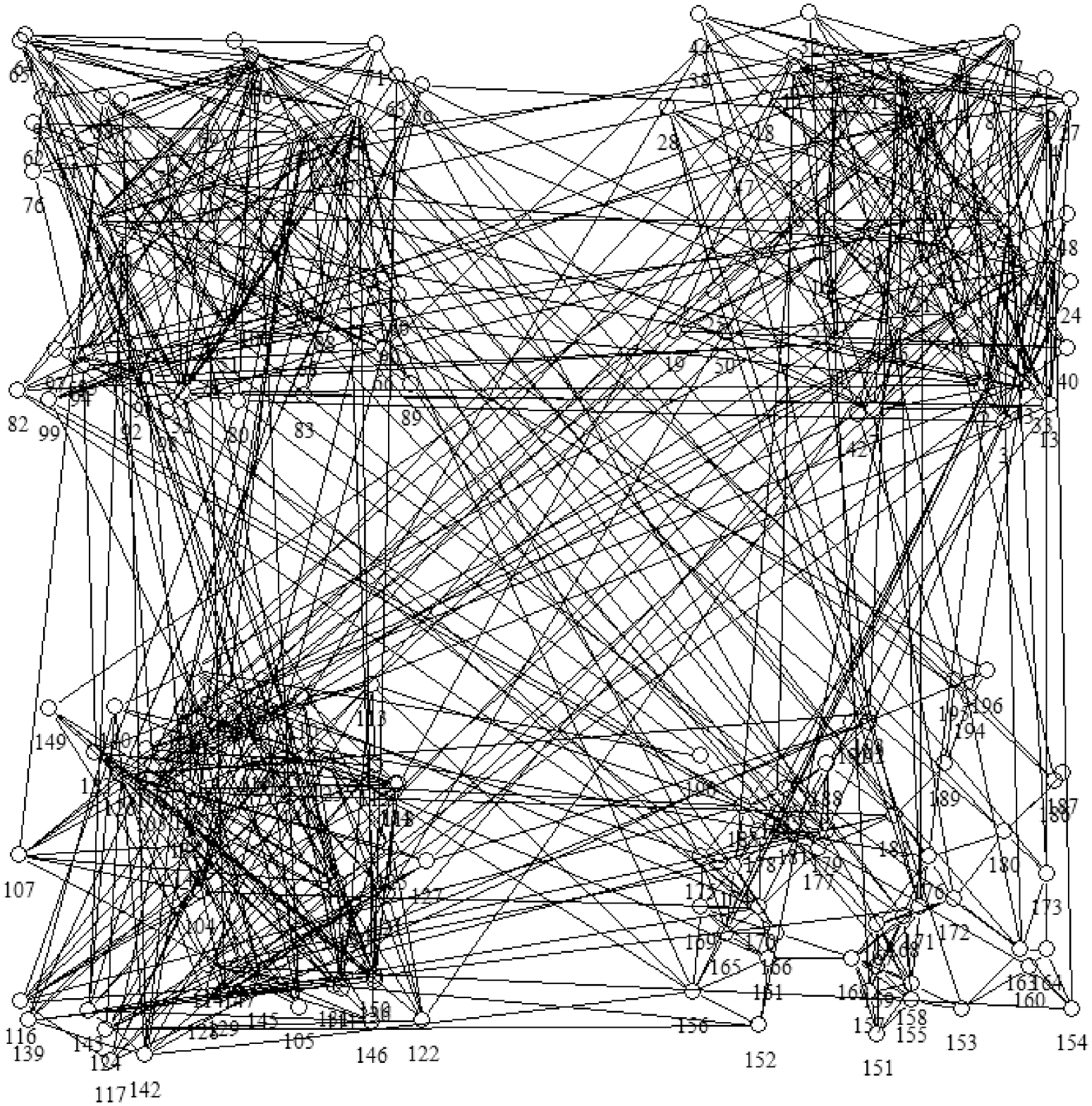} \\
  (a)   \hspace{9cm}  (b)   
  \caption{The two hybrid modular networks considered in this work.
           Each network has 4 communities (ER, BA, WS and GG as
           indicated in Fig~\ref{fig:ex_comm}), with intra-community
           degree equal to 6 and respective inter-community degrees 
           equal to 1 (a) and 2 (b).  These two networks are henceforth
           abbreviated as~\emph{Net 1} and~\emph{Net 2}. 
  }~\label{fig:two_nets} 
  \end{center}
\end{figure*}

In this section we illustrate how the modular equivalent model can be
used in order to better understand the transient confinement of
integrate-and-fire activation inside the communities.  In order to do
so, we consider two hybrid modular networks, each composed by ER, BA,
WS and GG modules with 50 nodes each (except GG, which has 46 nodes),
intra-community degree equal to 6 and inter-community degree equal to
1 and 2.  These two hybrid modular networks, henceforth abbreviated
as~\emph{Net 1} and~\emph{Net 2}, are shown in
Figure~\ref{fig:two_nets}. Observe that, because of finite-size effect
and statistical fluctuations, the GG networks have degrees smaller
than 6.

Let us choose node 1, which belongs to the ER community, as the
reference node.  The generalized hierarchical number of nodes
(i.e. the number of nodes inside each of the obtained equivalent
nodes) are given in Table~\ref{tab:n}.

\begin{table}
\centering
\begin{tabular}{|c|c||c|c|c|c|c|c|}  \hline  
               &  ER  &  1 &  5 & 28 & 16 &  0 &  0      \\  \cline{2-8}
     Net 1     &  BA  &  0 &  0 &  2 & 13 & 31 &  4      \\  \cline{2-8}
               &  WS  &  0 &  0 &  4 & 24 & 22 &  0      \\  \cline{2-8}
               &  GG  &  0 &  0 &  2 & 18 & 26 &  6      \\  \hline     
               &  ER  &  1 &  4 & 16 & 27 &  2 &  0      \\  \cline{2-8}
     Net 2     &  BA  &  0 &  0 &  1 & 10 & 31 &  8      \\  \cline{2-8}
               &  WS  &  0 &  0 &  0 & 15 & 34 &  1      \\  \cline{2-8}
               &  GG  &  0 &  0 &  4 & 29 & 13 &  0      \\  \hline
\end{tabular}
\caption{The number of nodes in the original networks which were
            associated to each of the
            equivalent nodes along the hierarchies 
            considering node 1 as the reference.
        }\label{tab:n}
\end{table}

\begin{figure}[htb]
  \vspace{0.3cm} 
  \begin{center}
  \includegraphics[width=0.8\linewidth]{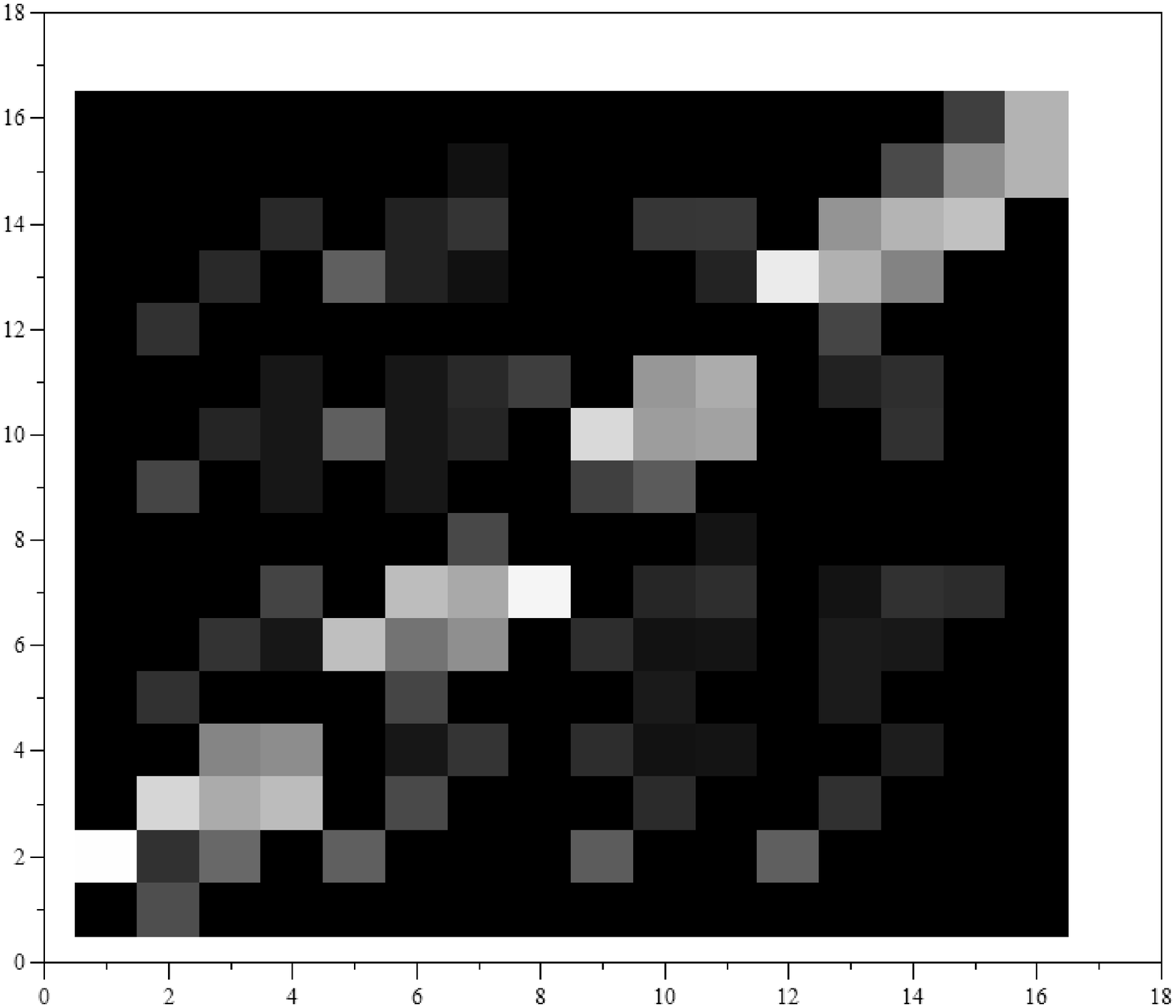} \\
  (a) \vspace{0.5cm} \\
  \includegraphics[width=0.8\linewidth]{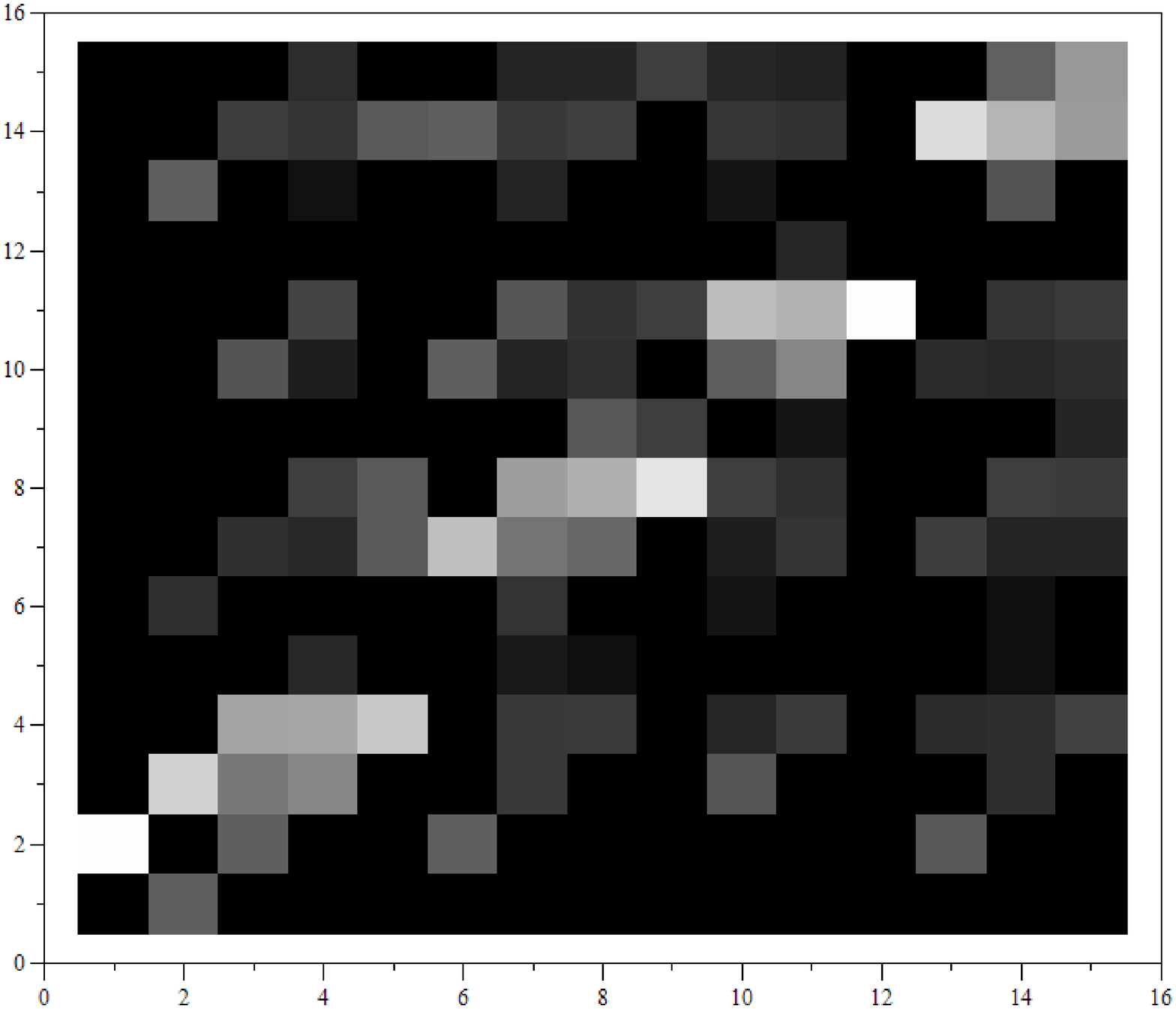} \\
  (b) 
  \caption{The weights of the equivalent models obtained for
                the first and second considered hybrid modular 
                networks, with node 1 as reference.  
                The gray-level intensities correspond
                to the weight values (white corresponds to maximum).
                The intensities are shown to their square root 
                values for the sake of better visualization.
  }~\label{fig:two_weights} 
  \end{center}
\end{figure}

The weights obtained for each of the two hybrid modular networks in
Figure~\ref{fig:two_nets} are respectively shown in
Figure~\ref{fig:two_weights}.  It is clear from this figure that
most of the activation transfers occur between the adjacent levels
(i.e. the weights near the main diagonal are higher). The second
network exhibits more intense off-diagonal weights as a consequence of
its higher inter-community degree (twice as much as for the first
network).  Such longer range connections are expected to intensify the
lack of simultaneity between the neurons belonging to each equivalent
node.

\begin{figure}[htb]
  \vspace{0.3cm} 
  \begin{center}
  \includegraphics[width=0.8\linewidth]{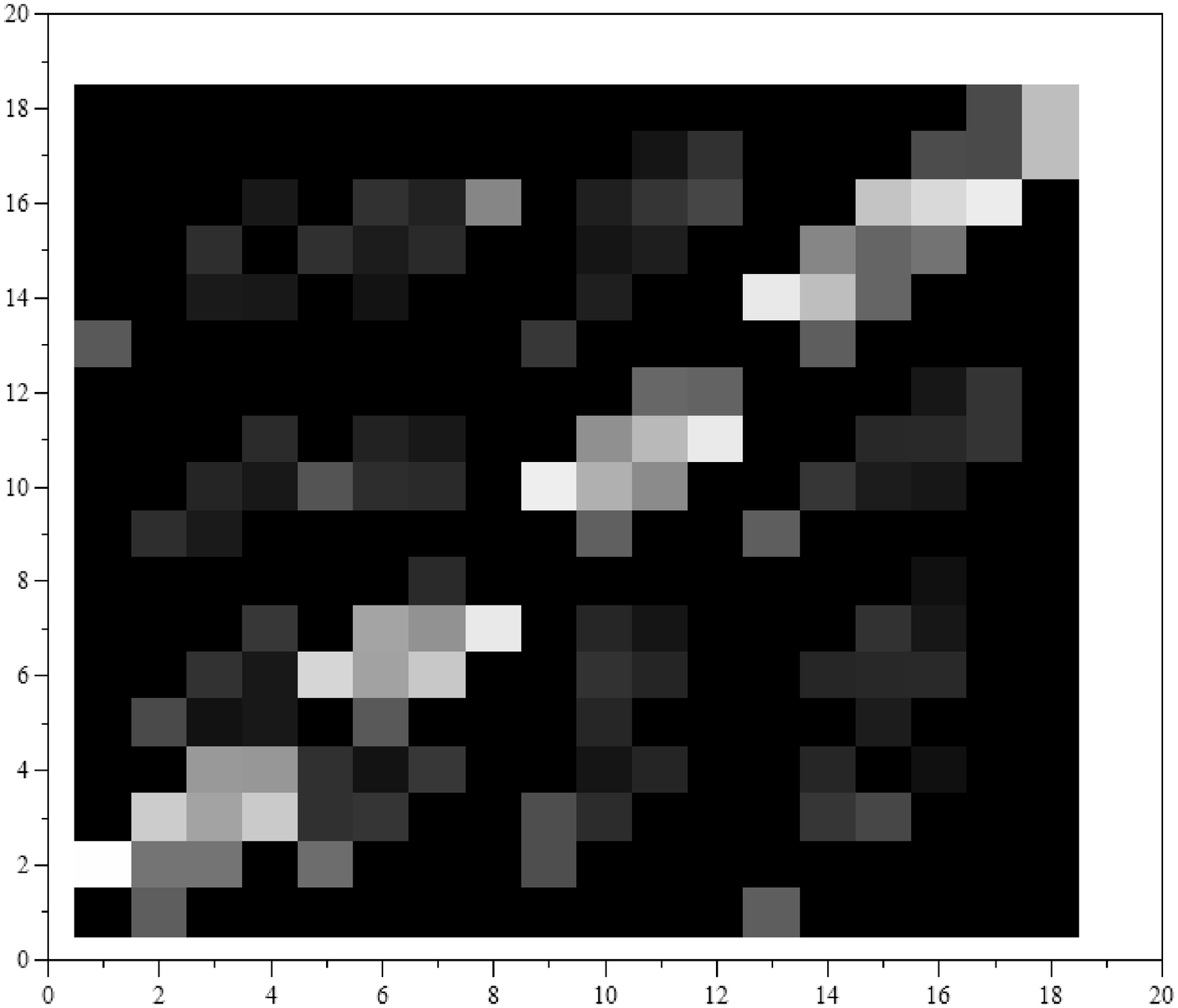} \\
  (a) \vspace{0.5cm} \\
  \includegraphics[width=0.8\linewidth]{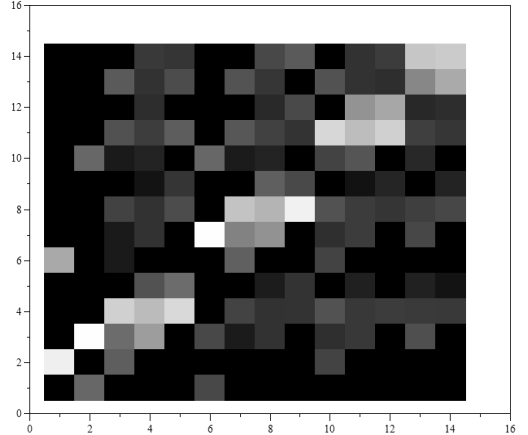} \\
  (b) 
  \caption{The weights of the equivalent models obtained for
                the first and second considered hybrid modular 
                networks, with node 50 as reference.  
          }~\label{fig:other_weights} 
  \end{center}
\end{figure}

Figure~\ref{fig:other_weights} shows the weights which are obtained by
selecting node 50 (also in the community ER) as the reference in Net 1
(Fig.~\ref{fig:other_weights}a) and Net 2
(Fig.~\ref{fig:other_weights}b).  While little changes can be observed
for Net 1 (with respect to the weights in Fig.~\ref{fig:two_weights}),
slightly more accentuated differences are noticed for Net 2.  Similar
trends have been obseved for other nodes, suggesting that different
choices of reference node have relatively little effect for networks
with smaller inter-community degree.

\begin{figure}[htb]
  \vspace{0.3cm} 
  \begin{center}
  \includegraphics[width=0.8\linewidth]{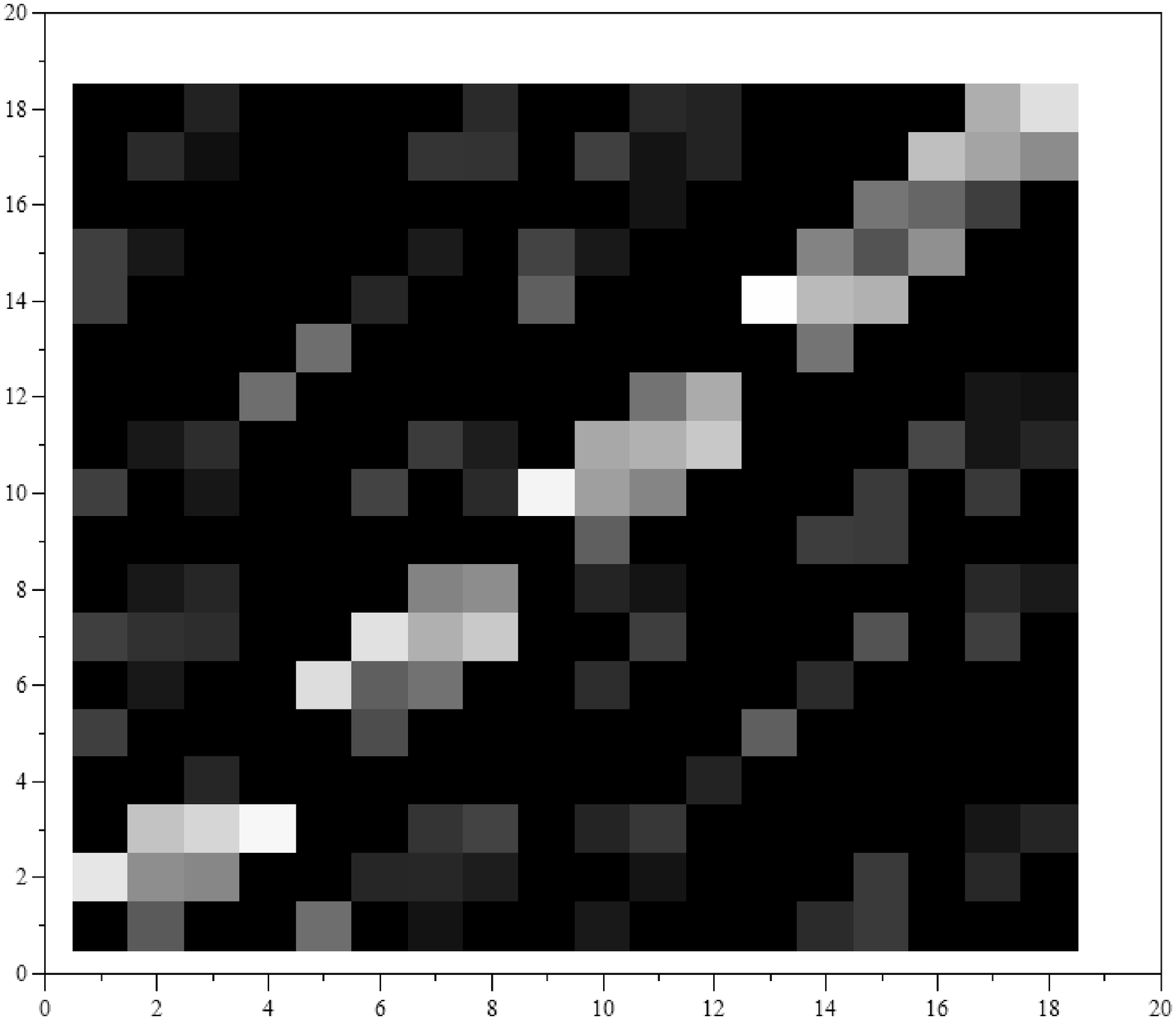} \\
  (a) \vspace{0.5cm} \\
  \includegraphics[width=0.8\linewidth]{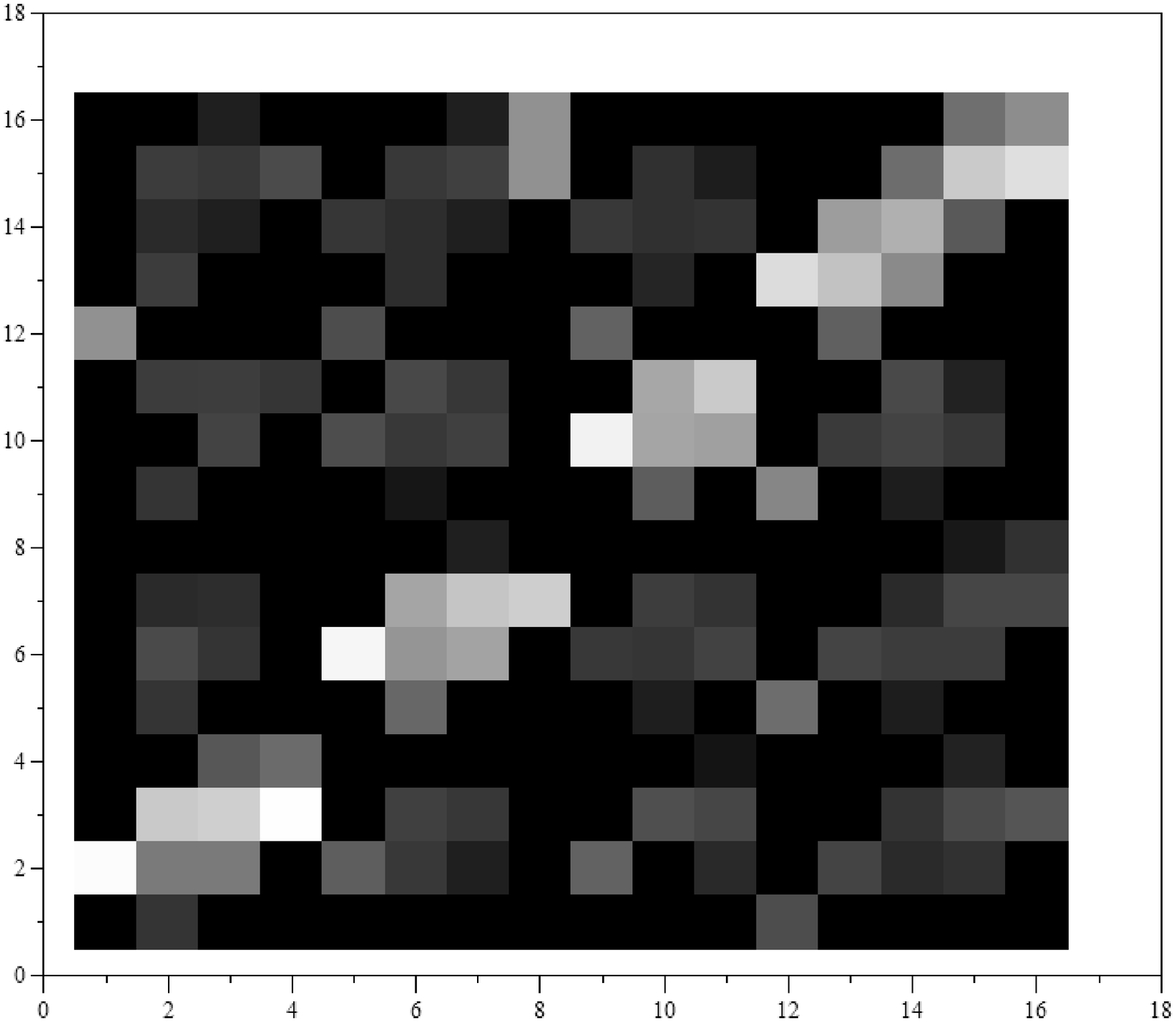} \\
  (b) 
  \caption{The weights of the equivalent models obtained for
                the first and second considered hybrid modular 
                networks, with node 170 (community WS) as reference.  
  }~\label{fig:other_comms} 
  \end{center}
\end{figure}

Figure~\ref{fig:other_comms} shows the weights which are obtained by
selecting node 170, now in the WS community as the reference node for
Net 1 (a) and Net 2 (b).  Interestingly, the weights implied by this
new reference node for Net 1 are similar to those obtained for node 50
as reference.  As in the previous example, larger differences of
weights were obtained for Net 2.

As discussed in~\cite{Costa_equiv:2008}, the timing of the main
avalanches in each community can be estimated by adding the number of
nodes from the concentric level 1 to the to the level exhibiting the
largest number of nodes inside each community.  Thus, for the ER
community, which has the maximum number of nodes in the third
hierarchy of Net 1, we expect the main avalanche to occur after nearly
24 time steps.  On the other hand, for the BA community, which has
maximum number of nodes at the fifth concentric level of Net 1, we
expect the main avalanche to take place after approximately
$1+5+28+16+2+13+4+24+2+18 = 113$ time steps.

\begin{figure*}[htb]
  \vspace{0.3cm} 
  \begin{center}
  \includegraphics[width=0.9\linewidth]{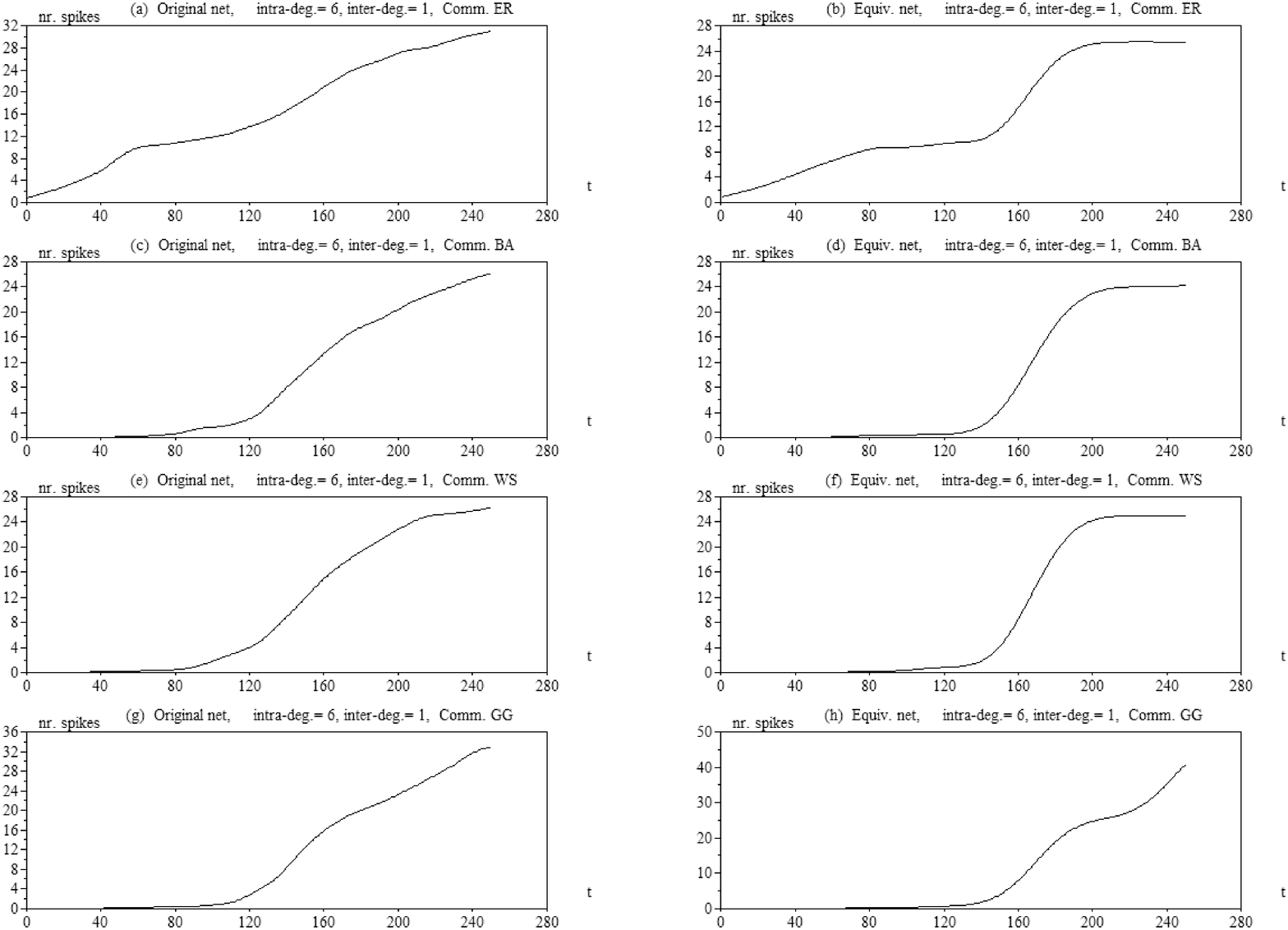} 
  \caption{The smoothed total number of spikes obtained for each
                community in Net 1 by using the original network
                (a, c, e, g) and respective equivalent models 
                (b, d, f, h).
  }~\label{fig:comp_net_1}
  \end{center}
\end{figure*}

Figure~\ref{fig:comp_net_1} shows the smoothed total number of spikes
obtained for each community in each of the original and equivalent
models of Net 1 in Figure~\ref{fig:two_nets}(a).  The total number of
spikes were smoothed through convolution with Gaussian functions with
standard deviations 8 (original net) and 16 (equivalent model) time
steps.  The different degrees of smoothing are necessary in order to
account for the fact that the equivalent model involves considerably
fewer neurons which spike simultaneously, leaving greater gaps along
the number of spikes curve.  Except for the plateaux in the curves
obtained for the modular equivalent models, an impressive agreement
can be verified between the results obtained considering the whole
network and the respective predictions of the dynamics by the modular
equivalent model.  Particularly precise estimations have been obtained
for the first community in both networks.  The adherence of the model
also increased with the inter-community degree, implying more accurate
predictions to be obtained for Net 2.  The timing of the respective
avalanches have been predicted with remarkable precision in all cases.

\begin{figure*}[htb]
  \vspace{0.3cm} 
  \begin{center}
  \includegraphics[width=0.9\linewidth]{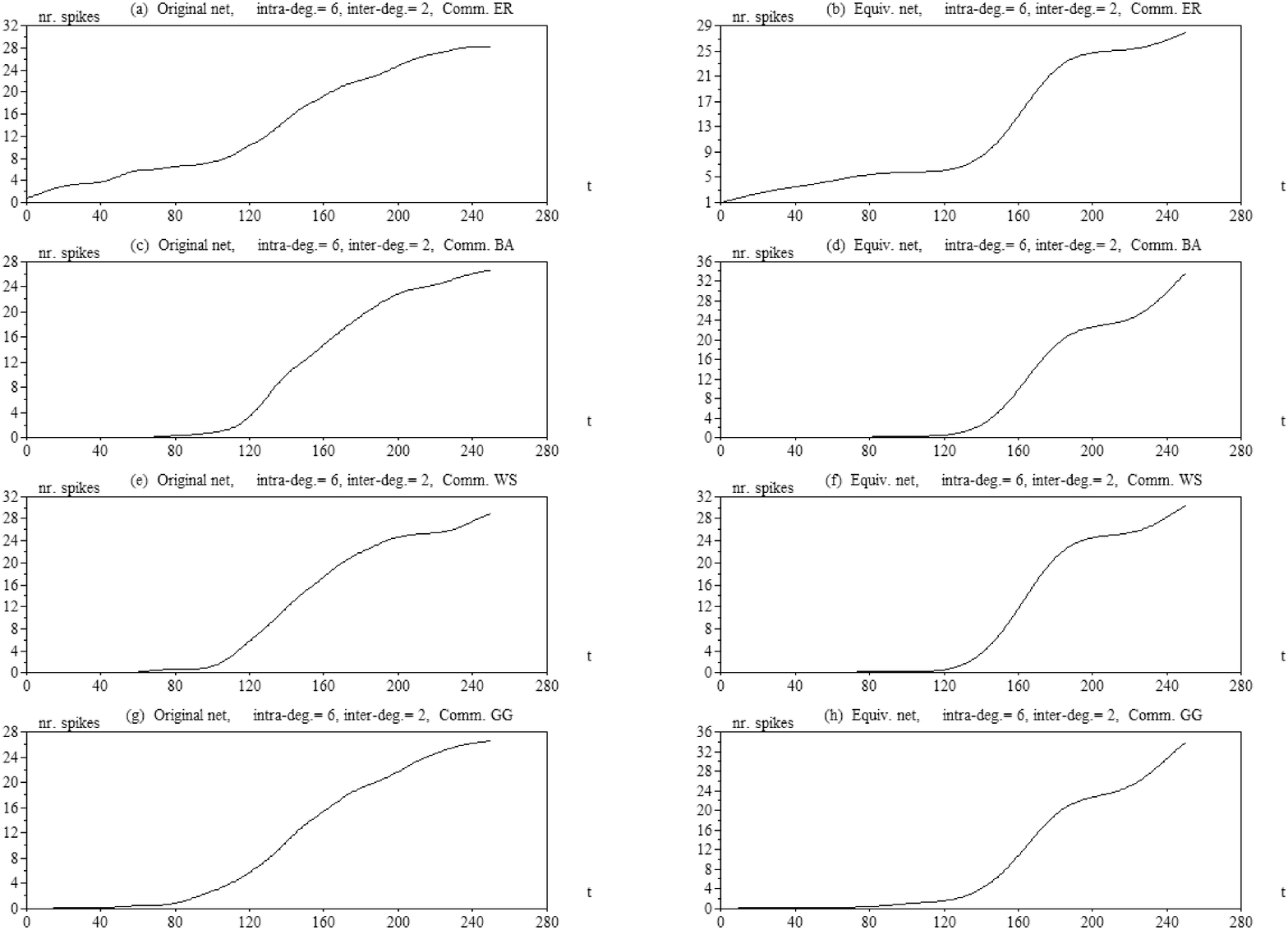} 
  \caption{The smoothed total number of spikes obtained for each
                community in Net 2 by using the original network
                (a, c, e, g) and respective equivalent models 
                (b, d, f, h).
  }~\label{fig:comp_net2}
  \end{center}
\end{figure*}

The impressive agreement between the results obtained by the modular
equivalent model, which incorporates only 30 nodes against the 196
nodes in the original networks, corroborates the fact that the
integrate-and-fire dynamics can be to its greatest extent captured by
considering the hierarchical organization of the original complex
neuronal networks.  Such a modeling and simulations make it clear that
the transient confinement of the spikes within the community where the
source is placed (this community is henceforth called the~\emph{source
community}) is a direct consequence of the integrate-and-fire
mechanism acting inside each equivalent node.  More specifically, the
source community activates first because most of the activation coming
from the source node goes into neurons belonging to that community.
This is an immediate consequence of the three following features: (i)
the source node sends, in the average, most of its axons towards nodes
in its own community; (ii) low off-diagonal weights of the modular
equivalent model are typically obtained for modular networks; and (ii)
longer times are required to induce nearly simultaneous spiking of
equivalent nodes related to a large number of nodes in the original
network. Indeed, the communities which do not contain the source node
will only start spiking after integrating for a long time the
typically the activation received from the source community.  This
time is particularly long because these communities initially receive
only a fraction of the activation in the source community, combined
with the large threshold required for these communities to become
activate, which typically takes place only after the main avalanche.
Indeed, the timing of the main avalanches obtained for the non-source
communities tended to be substantially larger than that of the source
community.

\begin{figure*}[htb]
  \vspace{0.3cm} 
  \begin{center}	
  \includegraphics[width=0.9\linewidth]{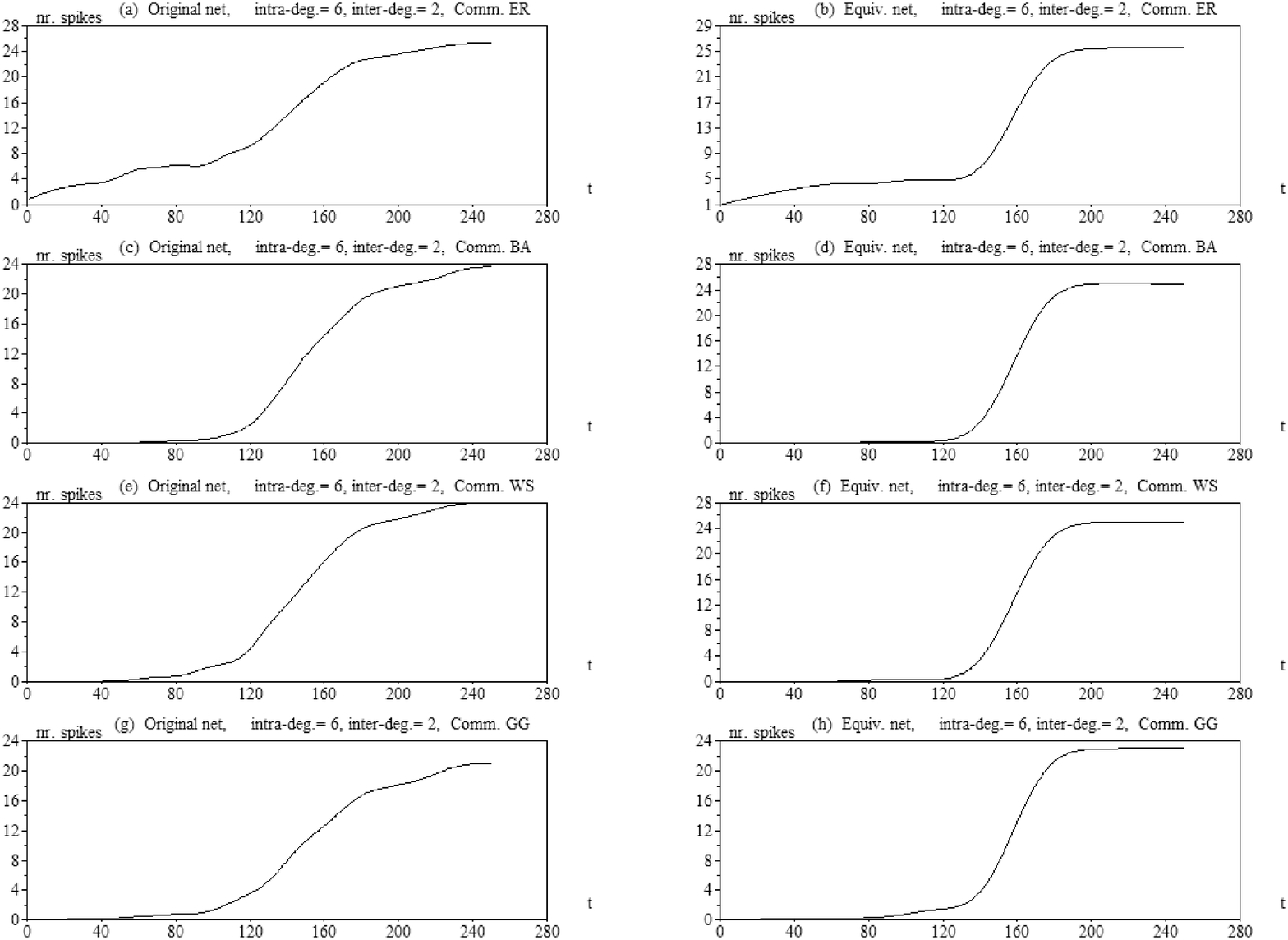} 
  \caption{The smoothed total number of spikes obtained for each
                community in Net 2 by using open versions of the
                 original network
                (a, c, e, g) and respective equivalent models 
                (b, d, f, h).
  }~\label{fig:comp_open_2}
  \end{center}
\end{figure*}

In order to learn more about the integrate-and-fire dynamics in
modular complex neuronal networks, it is interesting to compare the
total number of spikes obtained by comparing the dynamics obtained for
the whole original network and respective versions in which the
intra-ring connections have been removed~\cite{Costa:2004}.  Such
modified networks, henceforth called ~\emph{open networks}, can be
obtained by flooding the original network, starting from the reference
node, and removing all edges which are flooded in the same flooding
stage.  Figure~\ref{fig:comp_open_2} shows the smoothed total number
of spikes obtained for the open versions of Net 2 and its respective
modular equivalent model.  Interestingly, except for a small reduction
in the values of total number of spikes, the curves obtained for the
original network and respectively smoothed version are remarkably
similar.  Actually, a more careful comparison between the figures
reveals that the removal of the intra-ring connections implies a
slight increase in the derivative of the avalanche transition.  This
was indeed expected, but not to such a small degree, because the
intra-ring connections are known~\cite{Costa_equiv:2008} to disperse
the avalanche transition by implying the neurons related to each
equivalent node to spike in a less simultaneous fashion.

\section{Concluding Remarks}

Since its modern origins in the late 90's, and through its impressive
development ever since, the research area known as complex networks
has been characterized by special interest in the structural
properties of networks as well as with relationships with the
respectively obtained dynamics.  Because of the interesting structural
features implied by the presence of communities in networks, great
attention has been focused on studies of modular networks as well as
on community finding algorithms.  The present work lies at the heart
of the intersection of these two important issues in complex networks
research, namely the identification of communities and the interplay
between community structure and activation confinement, being also
related to fundamental aspects of neuroscience research.  More
specifically, in this work we have investigated in detail the
interesting effect of spiking confinement inside the communities of a
networks.  This has been accomplished with the help of an extension of
the chain equivalent model~\cite{Costa_equiv:2008}.  The main
contributions of the present work are listed and discussed as follows.

{\bf Incorporation of Intra-Ring Connections in the Chain Equivalent
Model:} When first proposed~\cite{Costa_equiv:2008}, the chain
equivalent model did not take into account the connections within the
same concentric level along the hierarchical organization of the
networks.  By incorporating self-connections at the equivalent nodes
of the chain model, it became possible to achieve a more precise model
of the integrate-and-fire dynamics unfolding along the hierarchical
structure of the complex neuronal networks.  The incorporation of the
intra-ring connections can lead to more precise estimation of the
original dynamics.  In addition, the role of the intra-ring edges in
undermining the simultaneity of the spiking at each level, as well as
the respective smoothing of the avalanche transitions, became clear
when these connections were removed.  Such an interesting effect
suggests that the removal of the intra-ring connections in the
original structure may lead to slight enhancements while mapping the
spiking patterns into clusters by means of the PCA
methodology~\cite{Costa_begin:2008, Costa_activ:2008}.

{\bf Smoothing of Activation Curves:} The curves of number of spikes
typically obtained for complex neuronal networks involve intense saw
oscillations~\cite{Costa_equiv:2008}, i.e. alternating high and low
number of spikes along subsequent time steps.  In the present work, we
have shown that the smoothing of the number of spike curves can yield
smoothed curves which are much more sensible and provide more
meaningful description of the integrate-and-fire dynamics in complex
neuronal networks.

{\bf Generalization of the Chain Equivalent Model for Asymmetric
Connectivity:} The chain equivalent model has been extended to
consider complex neuronal networks with asymmetric connections,
i.e. not every directed edge needs to have the respective
counter-directed edge as previously adopted.  Such a generalization
implied that while connections extending from the source node onwards
always extend from an equivalent node in one concentric level to an
equivalent node in the next level, backward connections can take place
between an equivalent level and equivalent nodes in any of the
previous levels.  Though the two hybrid modular networks considered in
this work were directed but had symmetric connections, such a
generalization was fundamental for allowing the extension of the
equivalent model to cope with modular complex neuronal networks.  

{\bf Development of the Modular Equivalent Model:} In this work, a
modular equivalent model was developed where the equivalent nodes are
determined with respect not only to the concentric levels, but also
taking into account the original communities.  The modular equivalent
model is a weighted complex neuronal network itself, suitable for
modeling any complex neuronal networks, directed/undirected, modular
or not, and with or without weights.  By considering two hybrid
modular networks, it has been presently shown that the modular
equivalent model, though containing only a handful of equivalent nodes
and connections, can predict with surprising accuracy the dynamics of
sophisticated complex neuronal networks, even in the presence of hubs
(BA) or lack of the small-world property (GG).  In addition, the
timing of the main avalanches in each community could be accurately
predicted by taking into account the number of nodes associated to the
concentric levels of the network.  Interestingly, different choices of
the reference node seem to have relatively little impact over the
weights of the equivalent model.

{\bf Removal of Intra-Ring Edges:} The contribution of the intra-ring
connections, namely the edges within the same concentric level, was
investigated by removing such nodes from the two considered hybrid
modular networks.  The respectively obtained results yielded similar
curves of total number of spikes along time, though with slightly
smaller values and with moderate increase of the derivative of the
avalanche transitions.  Such an effect suggests that the removal of
the intra-ring connections may contribute to improving the efficiency
of the integrate-and-fire methods for community identification
reported in~\cite{Costa_begin:2008,Costa_activ:2008}.

{\bf Full Understanding of the Confinement Effect:} The development of
the completely general modular equivalent model, as well as its
application to the modeling and accurate prediction of the
integrate-and-fire dynamics in two hybrid modular complex networks,
allowed a comprehensive understanding of the reason why the spiking
activation tends to remain confined, during a transient period of
time, inside the community to which the source node belongs.  Three
are the reasons leading to the confinement phenomenon: (i) the
tendency that most of the axons emanating from the source node are
sent to neurons in the same community; (ii) the fact that most of the
connections inside the source community are sent to nodes inside
itself; and (iii) the combined effect of the integrate-and-fire
dynamics, in the sense that the equivalent nodes associated to large
number of original nodes require much activation in order to reach the
respective threshold.  It should be observed that the increase of the
inter-community degree tends to reduce the confinement effect because
such an increase tends to undermine the effect in (iii).  Therefore,
such results confirmed the hypotheses advanced
in~\cite{Costa_begin:2008, Costa_activ:2008} about the origin and
properties of the confinement effect.

The relevance of the reported methodology and results is corroborated
by the several respectively implied prospects for future developments,
which are listed and briefly discussed as follows.

{\bf Dynamics After Removal of Activation:} It has been preliminary
verified that the removal of the external activation after a period of
time can contribute to emphasizing the identification of communities
by using the integrate-and-fire approach.  Such transient activations
may also lead to oscillation and pattern formation inside the complex
neuronal networks.  Further related investigations, facilitated by the
possibility of using the modular equivalent model in order to treat a
large number of networks in a reasonable period of time, are required
in order to fully assess the potential of such a strategy.

{\bf Networks with Decaying Activations:} As shown
in~\cite{Costa_activ:2008}, the incorporation of the biologically
realistic effect of activation decay with time (during
facilitation~\cite{Squire:2003}) can contribute to enhancing the
activation confinement inside the topological communities.  The
modular equivalent model can be immediately adapted to incorporate
such an additional dynamical feature, allowing the investigation of
the effect of the activation decay on the overall dynamics for a large
number of complex neuronal networks.

\begin{figure*}[htb]
  \vspace{0.3cm} 
  \begin{center}
  \includegraphics[width=1\linewidth]{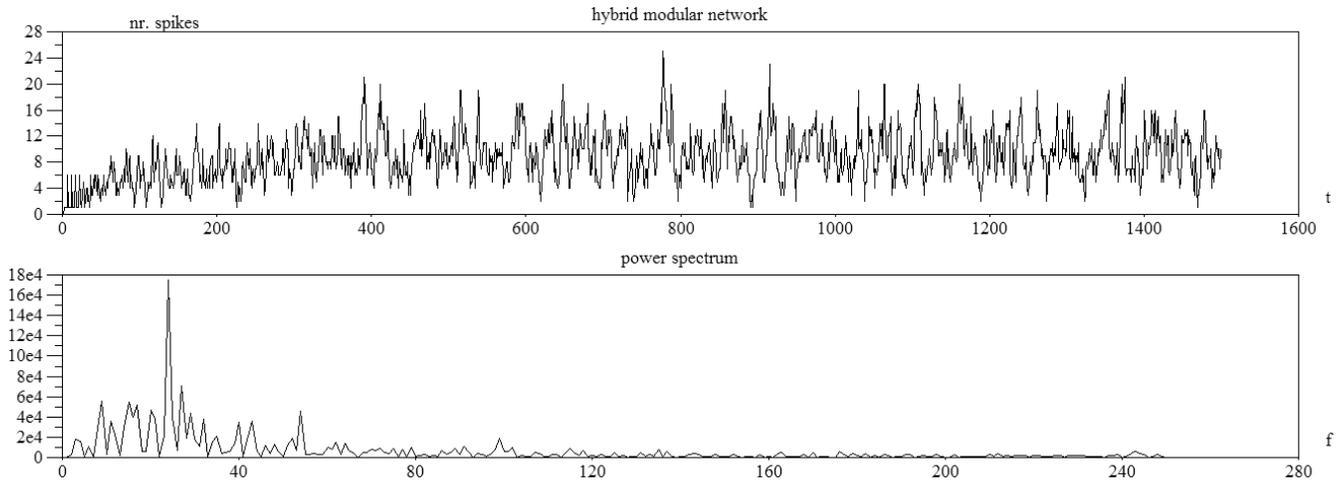} 
  \caption{The number of spikes (a) and power spectrum (b)
                obtained for limited integrate-and-fire
                activation of the hybrid network in Figure~\ref{fig:ex_comm}.
          }~\label{fig:power} 
  \end{center}
\end{figure*}

{\bf Networks with Limited Activation Transfers:} Another modification
of the integrate-and-fire dynamics which can be easily accommodated in
the modular equivalent model is the limitation of the output
activation.  More specifically, while in the present work we
considered the internally stored activation to be integrally
distributed among the outgoing axons, it is also possible to limit
such activations to a fixed value (the mostly stereotyped action
potentials~\cite{Squire:2003}).  Such a kind of dynamics is much more
biologically realistic than the currently considered model.  Results
obtained by considering this type of activation transfer limitation
are shown in Figure~\ref{fig:power}. Additional investigations
indicate that avalanches as well as the confinement effects occur even
when such a limitation is imposed and can be predicted by using a
respectively adapted modular equivalent model.  The limitation of the
activation transfer through the axons also led to moderate
improvements in community identification in several situations.

{\bf Equilibrium Dynamics:} Yet another promising future work would be
to characterize the equilibrium dynamics of the complex neuronal
networks after the activation source has been removed.  Particularly
worth of attention would be the identification of attractors of the
dynamics, as well as their relationship with the specific topological
properties and modularity of each complex neuronal networks.  It has
been preliminary observed that the equilibrium dynamics of complex
neuronal networks tend to exhibit oscillation patterns defined by
groups of nodes, not necessarily belonging to the same communities,
which acquire identical patterns of oscillation.  It would be
particularly interesting to relate such synchronization with the
topological and non-linear features of the integrate-and-fire complex
neuronal networks.

{\bf Activation-Induced Waves:} Figure~\ref{fig:power}(a) illustrates
the number of spikes obtained for a longer period of time (the
transfer of activation was limited to 1 at each axon in this case and
continuous external activation with intensity 1 was fed into the
system), with the transient dynamics taking place along approximately
the initial 500 time steps.  The respective power spectrum obtained by
considering the number of spikes from time 1001 to 1500, shown in
Figure~\ref{fig:power}(b), suggests that the number of spikes exhibits
a pronounced regular oscillation at frequency 52, corresponding to a
basic oscillation with period of 68 time steps in the original signal
in (a).  Observe also that the DC component (i.e. average of the
signal), not shown in the power spectra, has value close to the
amplitude of the waves.

This result, as well as other preliminary investigations, indicates
that the steady-state spiking of complex neuronal networks being
continuously fed, combined with limitation of the activation
transfers, tend to present such relatively well-defined oscillations.
Such oscillations may also be related to successive occurrences of
avalanches. Given that we have illustrated that the choice of the
source node seems to have a relatively little effect on the weights of
the equivalent model, such waves would have similar properties
irrespectively of the source position.  It would be interesting to
investigate to which point such waves induced by the
integrate-and-fire oscillations are possibly related to brain waves,
such as the alpha and beta rhythms, by considering the stimuli
activation.  In addition, how would the topological properties of
complex neuronal networks affect such induced waves?  Because the
intensity and timing of the avalanches are determined by the number of
nodes at the concentric levels, it is possible that the frequency of
the induced waves are defined by this hierarchical measurement.  Thus,
the progressive activation of the communities are expected to induce
variations of the wave frequencies.  Systematic additional
investigations are required in order to address such interesting
perspectives.

{\bf Analytical Extensions of the Modular Equivalent Model:} The
current work, as well as the previous investigation of the avalanches
phenomenon~\cite{Costa_equiv:2008}, were strongly based on the
development of progressively more general equivalent models, which
immediately allowed the identification of how the avalanches and
activation confinement arise in complex neuronal networks.  It would
be interesting to consider additional theoretical investigations,
especially based on non-linear systems theory, founded on the
simplified representation in terms of the equivalent models.

{\bf Comparison Between Several Integrate-and-Fire Methods for
Community Detection:} It has become clear that a series of
modifications of the integrate-and-fire dynamics can be implemented,
many of which have been found to contribute to improvements in
community finding.  It is now necessary to investigate systematically,
by considering several combinations of such modifications, which
particular parameters and integrate-and-fire specific dynamics tend to
yield the best community identification.

{\bf Comparison with other Community Finding Algorithms:} Though the
present study was not particularly concerned with the effectiveness of
the integrate-and-fire method for community identification, it would
be interesting to use the equivalent model, as well as the whole
networks, in order to investigate further how this method compares
with more traditional community finding algorithms.

{\bf Topology Arising as a Consequence of the Integrate-and-Fire
Dynamics:} So far, we have studied the dynamics given the topology of
the network.  The opposite is equally interesting, namely to change
the topology as a consequence of the dynamics.  Such investigations
are particularly relevant in neuroscience because they are
intrinsically associated to the all-important phenomenon of
\emph{memory}.  More specifically, it would be particularly exciting
to incorporate Hebbian-like mechanisms, where the connectivity would
be reinforced between nodes (or equivalent nodes) which tend to spike
at similar times. Would the communities be emphasized by such a kind
of topological dynamics?  How about the potential of such an approach
for pattern recognition, where patterns are trained by enhancing both
the intensity and number of the edges (each pattern would be trained
into a community)?  In particular, the identification of new patterns
to be trained, possibly requiring additional neurons or connections,
could be performed by taking into account the beginning activation
times of the neurons in each previous community.

{\bf Activation from Sets of Nodes:} Though attention has been
focused~\cite{Costa_nrn:2008, Costa_begin:2008, Costa_activ:2008,
Costa_equiv:2008} on the activation of the complex networks from
individual nodes, corresponding to the activation sources, it would be
interesting to consider the activation emanating from sets of nodes
placed or not within the same community.  Among the many related
possibilities, it would be especially worth investigating whether the
consideration of more than one source can lead to improved community
detection.  Another possibility would be to quantify the effect of
adding source nodes on the overall activation and spiking.

{\bf Applications to Real-World Networks:} Because of its theoretical
emphasis, only models of hybrid modular networks have been considered
in the present article.  This was especially critical because we
needed to know the communities \emph{a prior} in order to be able to
construct the modular equivalent models.  However, many interesting
studies can be obtained by considering real-world networks.  More
specifically, given a real-world network, its modular organization
could be identified by using a community finding algorithm (including
the integrate-and-fire method), allowing the construction of the
respective modular equivalent model, which would capture in a
simplified and explicit way the main aspects of the respective
non-linear dynamics, paving the way for further theoretical or applied
investigations.

{\bf Implications for Biology and Complex Systems:} For a long time
the `divide-and-conquer' strategy has been acknowledgee as a
particularly effective means for taming complexity.  The processing of
information in the brain constitutes no exception, involving spatial
and temporal modularization of the respective
processing~\cite{Zeki:1999, Hubel:2005}.  The close relationship
between the transient confinement of spiking within topological
modules in complex neuronal networks is intrinsically related to this
fundamental aspect of the brain organization.  Because of its
simplicity and efficiency in capturing the non-linear
integrate-and-fire dynamics, the modular equivalent model allows the
development of dynamic models of the functional architecture of the
brain by considering only the available interconnectivity between the
involved modules and information about the number of neurons and
processing layers in each of such modules.  Such a simplified but
effective model could be used to make all sorts of investigations and
predictions about the dynamics of brain activation, as well as to
study a number of neuronal diseases involving avalanches of brain
activation (e.g. Alzheimer's disease) and/or topological changes in
the connectivity.

\begin{acknowledgments}
Luciano da F. Costa thanks CNPq (308231/03-1) and FAPESP (05/00587-5)
for sponsorship.
\end{acknowledgments}

\bibliography{eqcomm}
\end{document}